\documentclass[final,5p,times,twocolumn,authoryear]{elsarticle}

\usepackage{braket,mathrsfs,mathtools,slashed} 
\usepackage[normalem]{ulem} 
\usepackage{amsmath} 
\usepackage{amsfonts}
\usepackage{dsfont} 
\usepackage{amssymb,fixmath,upgreek} 
\usepackage{soul} 
\usepackage{nccmath} 
\usepackage{enumitem} 
\usepackage{tikz}
\usepackage{tikzsymbols}
\usepackage{appendix}
\usepackage{lmodern} 
\usepackage{subfigure}
\usepackage{multirow}
\usepackage{hyperref}
\usepackage{cuted}
\usepackage{float} 

\definecolor{lime}{HTML}{A6CE39} 
\definecolor{darkpastelpurple}{rgb}{0.4, 0.1, 0.7}
\definecolor{darkgreen}{rgb}{.125,.5,.25} 

\allowdisplaybreaks 

\newcommand{\LambdaCDM}{{\textLambda\kern-0.06667em CDM} }

\newcommand{\dodoi}[1]{\href{https://doi.org/#1}{#1}}

\journal{High Energy Astrophysics}

\begin{document}

\begin{frontmatter}

\title{Measuring cosmic dipole with the GRB luminosity-time relation}

\author[first,1]{Jessica Santiago}
\affiliation[first]{organization={Aix Marseille Univ, Universite de Toulon, CNRS, CPT},
            city={Marseille},
            country={France}}
\affiliation[1]{organization={Leung Center for Cosmology \& Particle Astrophysics,
National Taiwan University},
            city={Taipei},
            country={Taiwan}}

\author[2]{Kerkyra Asvesta}
\affiliation[2]{organization={Aristotle University of Thessaloniki},city={Thessaloniki},
            country={Greece}}

\author[3,3b,3c,3d]{Maria Giovanna Dainotti}
\affiliation[3]{organization={National Astronomical Observatory of Japan, Mitaka},
            city={Tokyo},
            country={Japan}}
\affiliation[3b]{organization={The Graduate University for Advanced Studies, SOKENDAI},
            city={Kanagawa},
            country={Japan}}
\affiliation[3c]{organization={Space Science Institute, Boulder, CO 80301},
            country={USA}}
\affiliation[3d]{organization={Center for Astrophysics, University of Nevada, Las Vegas, NV 89154},country={USA}}

\author[1]{Pisin Chen}

\begin{abstract}
We present a new analysis of cosmic dipole anisotropy using gamma-ray bursts (GRBs) as high-redshift standardizable candles. GRBs are ideal probes for testing the cosmological principle thanks to their high luminosity, wide redshift range, and nearly isotropic sky coverage. For the first time, we employ the luminosity–time (L-T) relation, known in the literature as the bidimensional X-ray Dainotti relation, corrected for redshift evolution, to standardize a sample of 176 long GRBs detected by \textit{Swift}. We test for dipolar modulations in the GRB Hubble diagram using both the Dipole Fit Method and a new approach introduced here, the Anisotropic Residual Analysis Method. Both methods yield consistent results: a dipole amplitude of $A_d \simeq 0.6 \pm 0.2$ pointing towards (RA, DEC) $\approx (134^\circ \pm 30^{\circ}, –36^\circ \pm 21^{\circ})$ (equatorial coordinates). 
As shown in the Appendix, this corresponds to a boost velocity of the observer with respect to the GRB rest-frame in the antipodal direction from the dipole direction.
Extensive isotropy tests and 20,000 Monte Carlo simulations confirm that the detected signal cannot be explained by chance alignments or by the angular distribution of the GRB sample. We also show how, by incorporating a dipole term,  residual correlations are eliminated, showing that the dipole model provides a better fit than standard isotropic $\Lambda$CDM.
\end{abstract}

\end{frontmatter}

\section{Introduction}

The cosmological principle (CP) is one of the base assumptions of the current concordance $\Lambda$CDM model. It states that the universe, on large enough scales, has no preferred directions or locations. The $\Lambda$CDM model was born from merging the CP with observational evidence supporting a universe dominated by dark energy at late times.
Besides its remarkable success, growing tensions -- especially the $>5\sigma$ Hubble constant discrepancy -- challenges its validity~\citep{dainotti2021ApJ, Dainotti2022Galax..10...24D,Aluri:2022hzs, DiValentino:2022fjm, Abdalla:2022yfr, Perivolaropoulos:2021jda, Buchert:2015wwr, Hu:2023jqc, Murakami:2023,Dainotti2025JHEAp..4800405D}.

Furthermore, claims regarding strong deviations from large-scale homogeneity and isotropy provide additional concerns on $\Lambda$CDM's structural basis \citep{Colin2011, Migkas:2021zdo, Sah:2024csa, peerywatkins2018, watkins2023, howlett2022}. 
While small-scale structures are expected to cause inhomogeneities~\citep{Qin:2021tak, Kalbouneh:2024}, it is generally assumed that statistical isotropy and homogeneity emerge beyond $100h^{-1}$Mpc ~\citep{2005MNRAS.364..601Y, Laurent:2016eqo, Labini_2011}. 
The Cosmic Microwave Background (CMB) provides the strongest observational evidence on that respect, revealing a remarkable isotropic distribution on small angular scales. 
Its dipole, interpreted as kinematic due to the Solar System’s motion at $369\;\text{km s}^{-1}$, points towards $(l, b) = (264.021 \pm 0.011, 48.253 \pm 0.005)$ (galactic coordinates) \citep{Planck:2018nkj}. 

A key test of the cosmological principle (CP) lies in determining whether the CMB dipole is entirely of kinematic origin. A way to probe this is by measuring our velocity relative to the so-called matter-frame. If the CP holds, this velocity must match our motion relative to the CMB~\citep{Maartens:2023tib}. 
In a number of recent studies~\citep{Sorrenti:2022zat, Singal:2011dy, rubart2013, Dam:2022wwh, Secrest:2020has} the obtained velocity has failed to recover the CMB results, highlighting the need of additional investigation.

To further test for the presence of large-scale anisotropies, wide-ranging observational probes are required.
At low to intermediate redshifts, Supernovae Type Ia (SNe Ia) studies have given us contradicting results regarding the statistical significance of the measured isotropy deviations \citep{2019A&A...631L..13C, Cai_2012, Rahman:2021mti, Krishnan:2021jmh, Hu:2024qnx}. 
At high redshifts, analysis using the Ellis \& Baldwin test \citep{ellisbald1984} report dipole amplitudes significantly higher than the kinematic CMB dipole -- some exceeding $4\sigma$ significance~\citep{Singal:2011dy, rubart2013, Bengaly:2017slg, Secrest:2020has, Secrest:2022uvx, Colin:2017juj, Wagenveld:2023kvi, Land-Strykowski:2025gkz} -- though these claims remain contested \citep{2012MNRAS.427.1994G, Tiwari:2015tba, Siewert:2020krp}.

Aiming to further probe the high-redshift regime, gamma-ray bursts (GRBs) have an encouraging set of properties which makes them an ideal candidate to test the CP. 
GRBs are the brightest cosmological sources, with isotropic luminosities of $10^{48} - 10^{54} \;\text{erg s}^{-1}$ and average sample redshift around $z\sim2$. Their sky coverage exceeds 70\% for missions like BATSE, Fermi, Swift, and Konus-Wind. Moreover, well-studied emission properties have revealed standardizable relations, enabling their use as standard candles, see ~\cite{delvecchio16,Dainotti2013a,dainotti2022opticaldata,amati2006,yonetoku2004}. 
In this work, we focus on the $\log{L_a}-\log{T_a^*}$ relation (L-T or Dainotti relation) \citep{Dainotti2008}.

The paper goes as follows: Section \ref{S: data} describes the data and sample selection. Section \ref{S: standardizing} covers the standardizing L-T relation and the evolution corrections used. Section \ref{S: dipole} briefly reviews the \emph{Dipole Fitting Method} and introduces our new approach, the \emph{Anisotropic Residual Analysis Method}. Results are presented in Section \ref{S: results}, followed by a discussion, comparison with related studies and conclusion in Section \ref{S: conclusion}.

\begin{figure*}
\centering
\subfigure{
  \includegraphics[width=0.62\textwidth]{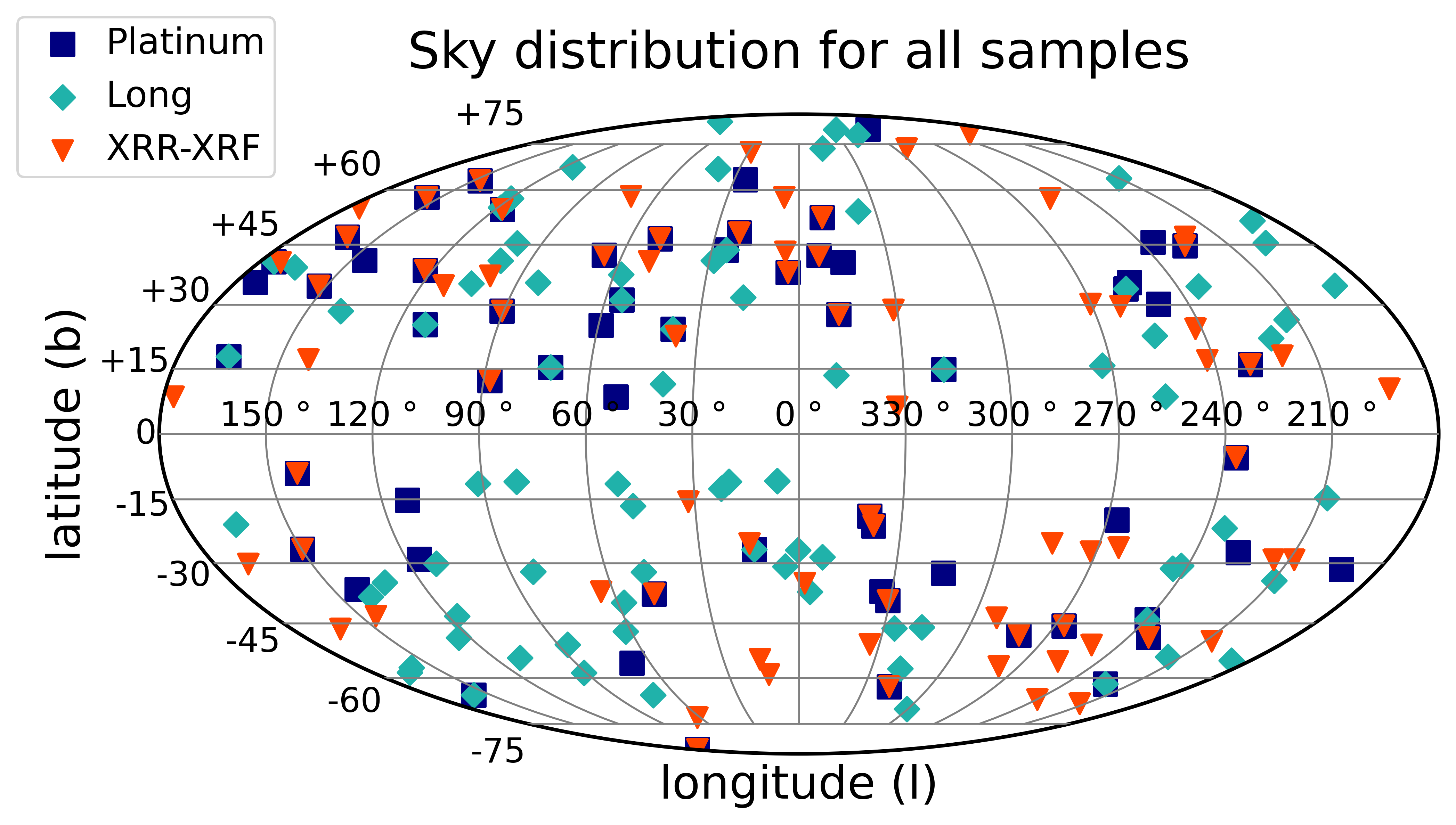}
}
\hfill
\subfigure{
  \includegraphics[width=0.32\textwidth]{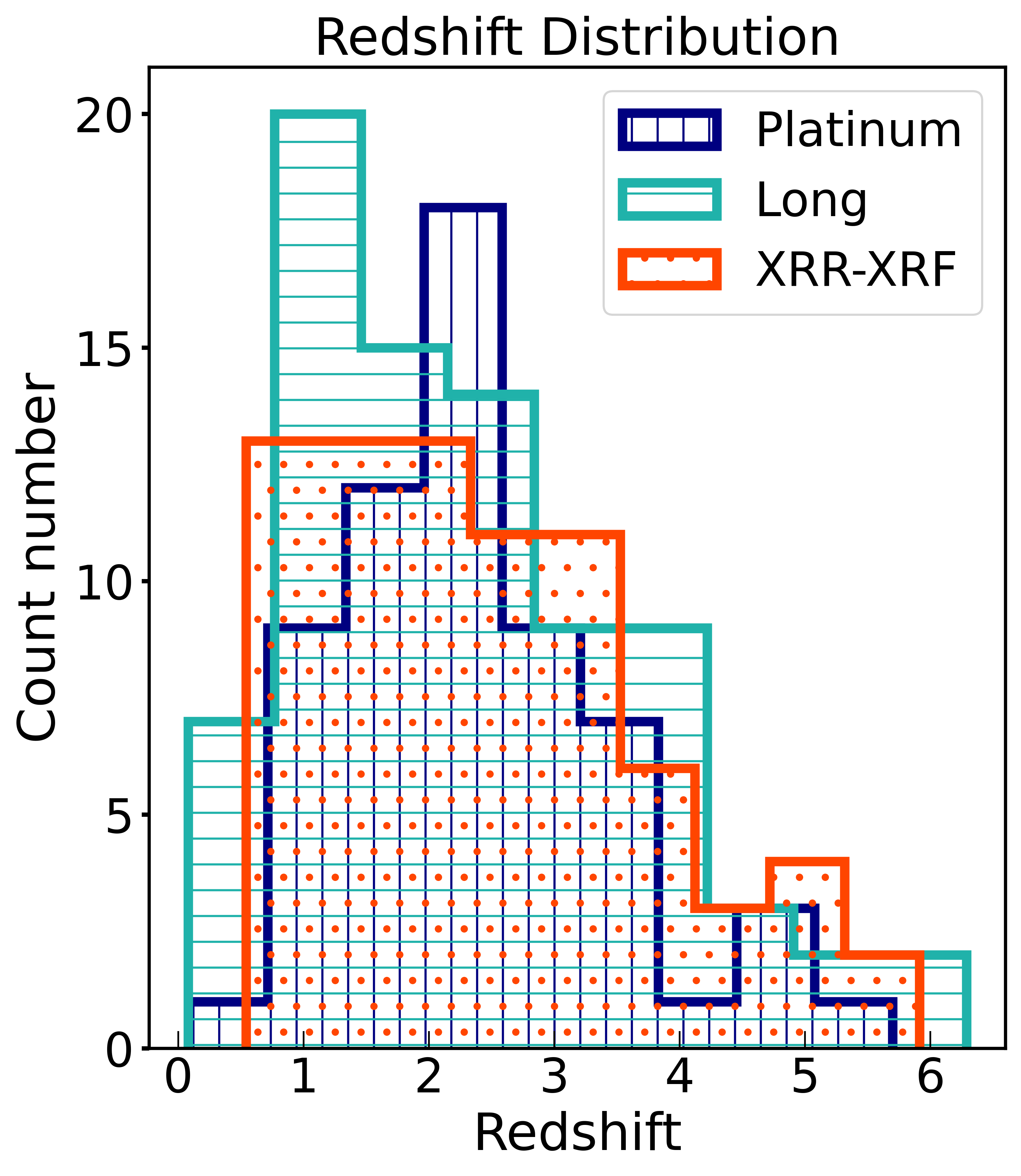}
}
\caption{Sky distribution in galactic coordinates (left) and redshift distribution (right) for individual data samples.}
\label{fig:wide_subfig}
\end{figure*}

\section{Data}
\label{S: data}
We used \textit{Neil Gehrels Swift Observatory} data (\textit{Swift}). 
The catalog consists of 255 GRB sources, all classified as `long' GRBs with $T_{90}>2$~seconds, where $T_{90}$ is the time interval during which $90\%$ of the burst’s total observed gamma-ray counts are detected. 
Besides all 255 GRBs being classified as `long', sub-samples can still be defined in the following way:
\begin{itemize}
    \item Platinum (61 GRBs): GRBs that obey a strict selection criteria, with low scatter in the fundamental plane (see \citep{Dainotti:2020azn} for details);
    \item X-ray flashes (XRFs) and X-ray rich GRBs (XRR) compose a sample of 76 GRBs.  These are softer and often less energetic, with emission peaks on the X-ray range;
    \item SN GRBs associated with SNe Ib/c for which the SNe Ib/c has been detected (21 GRBs).
    \item Long GRBs (81 GRBs): all the remaining long GRBs which do not fit into any of the previous classes.
    \item Combined GRBs (176 GRBs): all the above classes combined, with duplicates removed. Duplicates are present since many GRBs in the Platinum sample belong to either Long, XRR-XRFs or SN classes as well.
\end{itemize}
The Platinum sample used is an extended sample compared to \citep{dainotti2020a}. 

\subsection{Sample Selection}
Observationally, a GRB is composed of a \textit{prompt emission} observed mainly in $\gamma$-rays and hard X-rays, and an afterglow emission. In about 60\% of the cases the afterglow presents a plateau, a relatively flat part of the light curves following the decay of the prompt emission \citep{Srinivasaragavan2020}.

Aiming to obtain a reliable set of standardizable GRBs, we performed a selection cut based on the following criteria: 
\begin{enumerate}
    \item $\Delta F_{peak} < F_{peak}$\;;
    \item $\Delta K_{a} < K_{a}$\;;
    \item $\Delta K_{p} < K_{p}$\;,
\end{enumerate}
where $\Delta X$ represents the uncertainties on the variable $X$. Thus, $\Delta F_{peak}$ is the uncertainty on the peak flux, while $\Delta K_a$ and $\Delta K_{peak}$ are the uncertainties on the K-correction for the plateau emission and the prompt emission respectively \citep{Bloom2001}.
These conditions ensure that the relative error on these variables is $<$ 1. After applying these criteria, the final number of GRBs in each class is:
Platinum (61), XRR-XRF (76), Long (81) and Combined (176).
Figure~\ref{fig:wide_subfig} shows the Mollweide projection (left panel) and redshift distribution (right panel) for the individual samples.
The Combined sample covers a redshift range $0.036 <z< 6.30$,  with a mean redshift $\Bar{z} = 2.4$. A more thorough analysis of the Combined Sample angular distribution isotropy can be found in Appendix \ref{A: Platinum sample}.
All the observed quantities are measured in the heliocentric reference frame. 

\section{Standardizing GRBs with the Dainotti relation}
\label{S: standardizing}
\begin{figure}
\subfigure[$\log L_a$ vs. $\log T_a^*$ linear relation]{
  \includegraphics[width=0.435\textwidth]{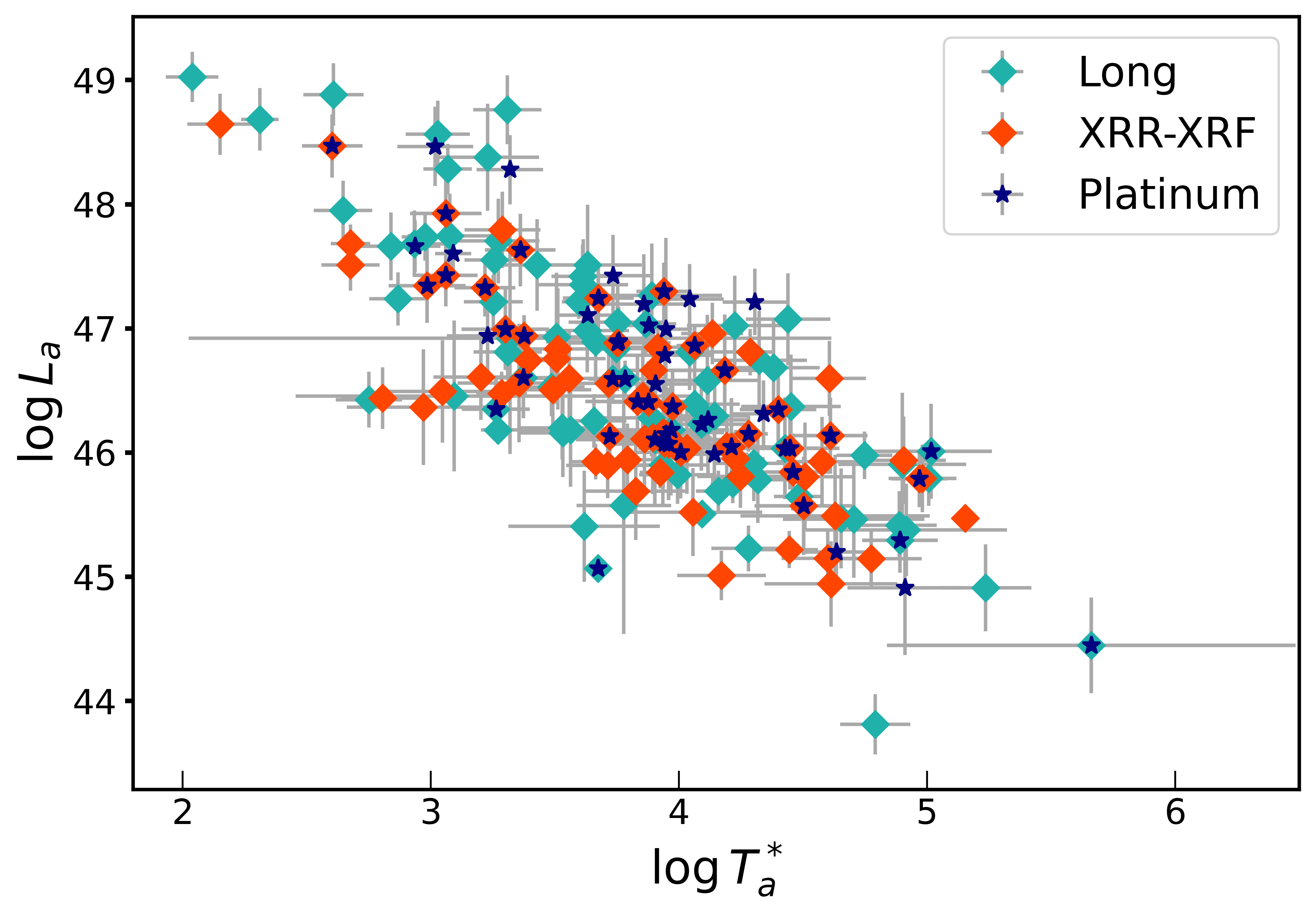}
  \label{fig:La-Ta}
}
\hfill
\subfigure[Best-fit posterior distributions]{
  \includegraphics[width=0.435\textwidth]{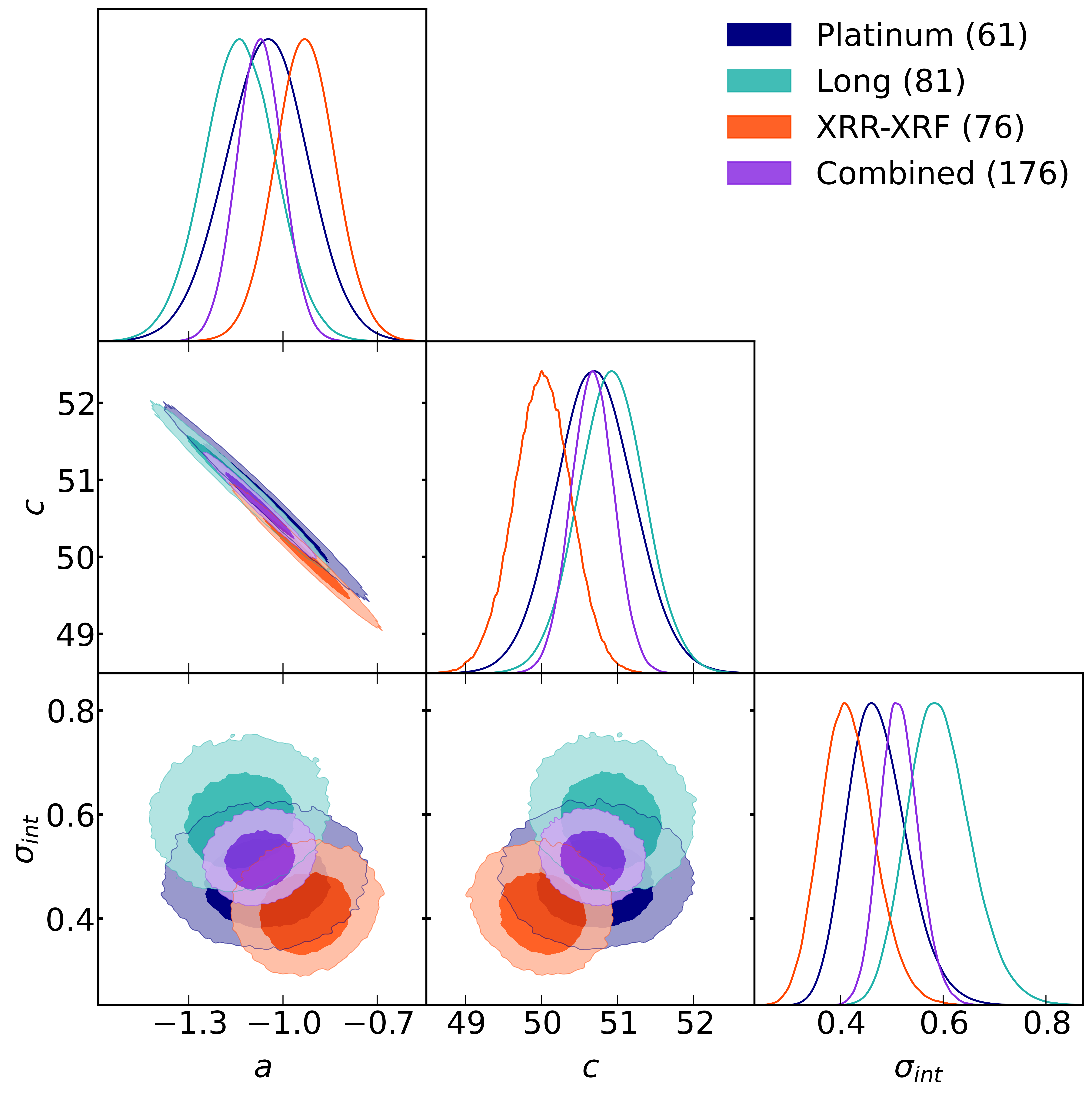}
  \label{fig:CornerPlot2D}
}
\hfill
\subfigure[Hubble Diagram for Individual GRB samples]{
  \includegraphics[width=0.435\textwidth]{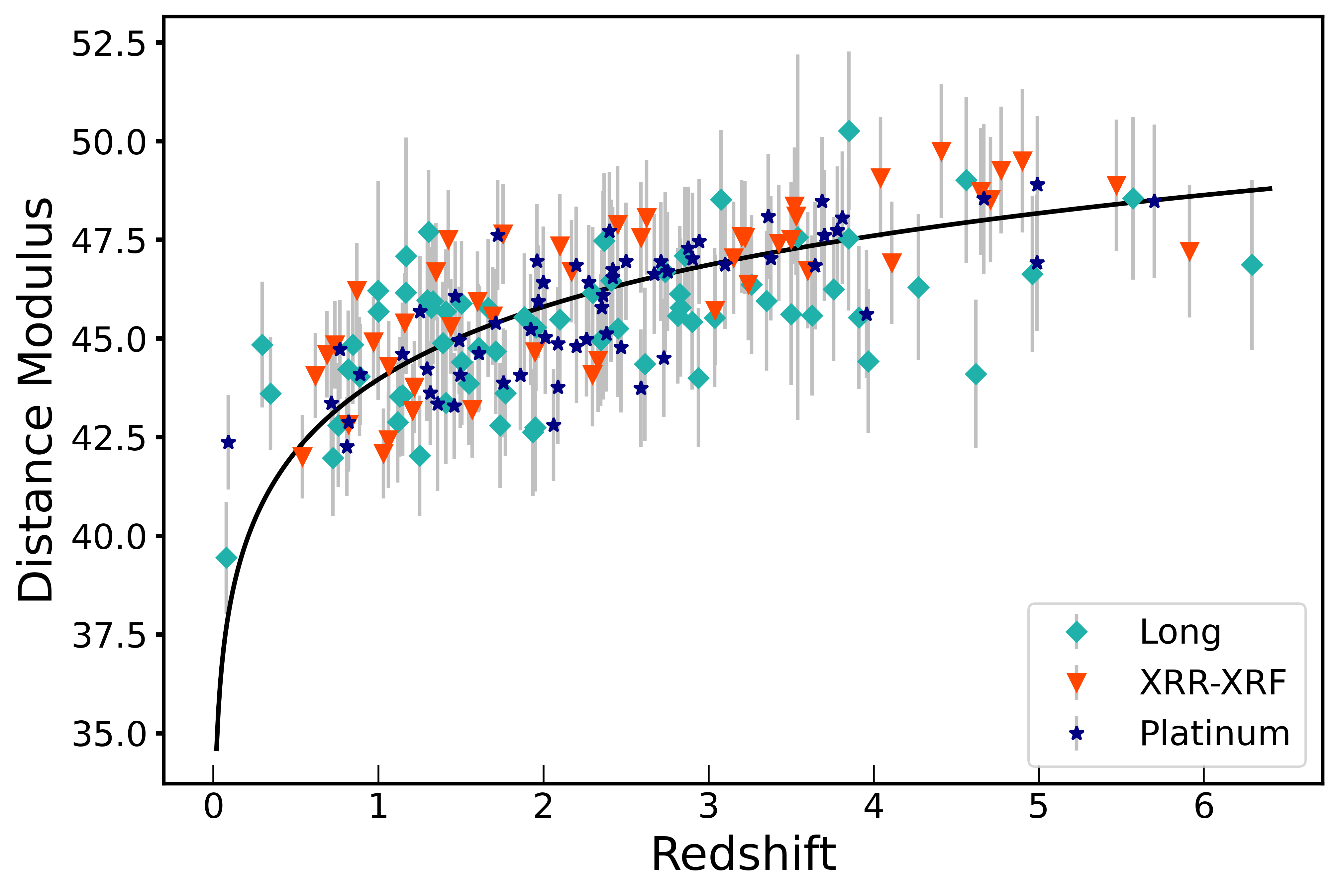}
  \label{fig:Hubble2D}
}
\caption{(a) The L-T diagram showing the sample correlation, (b) best-fit standardizing parameters and (c) Hubble diagram applying the best-fit values and redshift evolution.
In case of duplicates among samples, 
only the \emph{Platinum} value is displayed.}
\label{fig:multi2D}
\end{figure}

The Dainotti relation is an empirical and linear relation \citep{Dainotti2008} among the logarithms of the X-ray plateau luminosity $L_a$ and its rest-frame duration $T_a$. It reads as:
\begin{equation}
    \log L_a = \log (4 \pi D_{L}^2 F_{a} K_{a})=a \log T_{a} + c\;.
    \label{eq:2Drelation}
\end{equation}
The coefficients $a$ (slope), and $c$ (normalization parameter) are determined together with the intrinsic scatter $\sigma_{int}$, $F_{a}$ is the measured X-ray energy flux at the end time (rest frame) of the plateau emission $T_{a}$, $K_a$ is the K-correction and $D_L$ is the luminosity distance.

The intrinsic scatter is an additional dispersion in the observed correlation beyond measurement uncertainties.
It reflects a hidden source of uncertainties not captured by the variable uncertainties. 

Another important aspect is the possible redshift evolution of intrinsic correlations. \cite{Dainotti2013a} showed that the L-T relation is redshift-dependent if not corrected, potentially biasing cosmological inference. Evolution may arise from selection effects, instrumental limits, or true changes in GRB progenitors. This is corrected using the Efron–Petrosian (EP) method \citep{Efron1992,Dainotti2013b}. The evolution-corrected variables are:
\begin{align}
    F_{a_{cor}} = \frac{F_a\; K_a}{(1+z)^{k_{L_a}}}\;\;\;
    \text{and} \quad\;T_a^{*} = \frac{T_a}{(1+z)^{k_{T_a}}}\;,
    \label{eq:evolution}
\end{align}
with evolutionary parameters $k_{T_a} = -1.25 \pm 0.275$ and $k_{L_a} = 2.42 \pm 0.575$, obtained from \cite{Dainotti:2022ked}.

Figure \ref{fig:La-Ta} shows the evolution-corrected L-T relation for the Long, XRR-XRF and Platinum sample. We also performed a Pearson test to measure the linear correlation $r$ between $\log L_a$ and $\log T_a^*$.
The corresponding $p$-value estimates the likelihood of obtaining $r$ by chance.
The obtained $r$ and p-values are:
\begin{itemize}[leftmargin=*,label=$\;$]
   \item \textit{Platinum:} $r = -0.80$, $p-$value $= 2\times10^{-14}$;
    \item \textit{Long:} $r = -0.77$, $p-$value $= 4\times10^{-17}$;
     \item \textit{XRR-XRF:} $r = -0.76$, $p-$value $= 2\times10^{-15}$; 
      \item \textit{Combined:} $r = -0.76$, $p-$value $= 2\times10^{-34}$. 
\end{itemize}
The second equality of equation \eqref{eq:2Drelation} defines a luminosity distance for GRBs as:
\begin{equation}
    \log D_L = \frac{a}{2} \log T^*_{a} -\frac{1}{2}\log F_{a_{cor}} +\frac{(c - \log 4\pi)}{2}  \;.
    \label{eq:DL2D}
\end{equation}
From $D_L$, one can obtain the GRB distance modulus:
\begin{equation}
    \mu_\text{grb} = \frac{5}{2} \;(a\log T^*_{a} - \log F_{a_{cor}} + c -\log_{10}(4\pi)) +25\;.
    \label{eq:mu grb 2D}
\end{equation}
Figures \ref{fig:CornerPlot2D} show the joint and marginal posterior distributions for the parameters $(a,c,\sigma_{int})$ and \ref{fig:Hubble2D} the Hubble diagram obtained using the best-fit values for the L-T relation.

\section{Dipole Measurement Methods}
\label{S: dipole}

Several methods exist for detecting dipolar anisotropies in the matter distribution. In this work, we apply both the established \emph{Dipole Fit Method}~\citep{Mariano:2012wx} and introduce a new approach: the \emph{Anisotropic Residual Analysis Method}.

\subsection{Dipole Fitting Method}
This method involves directly fitting observational data to a dipole-modulated cosmological model. 
Specifically, we test for a dipole in the GRB distance modulus by adding a dipole term to the standard $\Lambda$CDM prediction:
\begin{equation}
    \mu_{\text{dip}}(z) = \mu_{\Lambda CDM}(z) + \mathbf{A_d}\cdot\hat{\mathbf{n}}_i = \mu_{\Lambda CDM}(z) + A_d\cos\vartheta_i\;, \label{eq: mu dipole}
\end{equation}
where $\mu_{\Lambda CDM}(z)$ corresponds to the theoretical distance modulus assuming a flat $\Lambda$CDM universe with fiducial values given by $H_0 = 73.04$ Km/sec/Mpc and $\Omega_m = 0.334$. The addition of a dipole vector $\mathbf{A_d}$ in the distance modulus introduces three new degrees of freedom,  given by the dipole amplitude $A_d$ and its direction $\vec{d}$, in our analysis determined by its right ascension (RA) and declination (DEC). Here, $\hat{\mathbf{n}}_i$ is a unitary vector pointing towards the $i$th GRB's direction and $\vartheta_i$ is the angle between the direction of the dipole $\vec{d}$ and $\hat{\mathbf{n}}_i$, calculated as:

\begin{equation}
    \cos\vartheta_i = \cos \delta^i_{grb}\cos \delta_{d} \cos(r^i_{grb} - r_{d})+\sin \delta^i_{grb} \sin \delta_{d}\;,
    \label{eq:vartheta2}
\end{equation}
where $(r^i_{grb}, \delta^i_{grb})$ are the directions of the $i$th GRB on the celestial sphere and correspond to the right ascension and declination in degrees. The pair $(r_{d}, \delta_{d})$ represents the Equatorial coordinate directions of the dipole accordingly.

Using \eqref{eq:mu grb 2D} as the observed GRB distance modulus, we perform a Bayesian analysis via MCMC to estimate the joint posterior distributions of the standardization L-T relation parameters $(a, c, \sigma_{\text{int}})$ and the dipole parameters $(A_d, \text{RA}, \text{DEC})$. The corresponding $\chi^2$ is computed by comparing the theoretical distance modulus \eqref{eq: mu dipole} to the observed values:
\begin{equation}
    \chi^2_\text{grb} = \sum_N \left(\frac{\mu_\text{dip} - \mu_\text{grb}}{\sigma_{\mu_{obs}}}\right)^2\;,
\end{equation}
where $\sigma_{\mu_{grb}}$ is the error in the observed distance modulus:
\begin{equation}
    \sigma_{\mu_{grb}} = \frac{5}{2}\sqrt{(a \;\sigma_{\log T_a})^2  + (\sigma_{\log F_{a_{cor}}})^2  +  4\;\sigma_{int}^2.}
    \label{eq:sigma_mu}
\end{equation}
The results obtained with this method are presented in Section \ref{S: results}.

\subsection{Anisotropic Residual Analysis Method}
\label{S: Residual Analysis theory}
In this section we introduce a novel dipole ana-lysis method. It relies on the premise that if a dataset exhibits an underlying directional dependence--such as a dipolar modulation--by fitting it with a model that neglects its angular structure will cause the anisotropic signature to manifest in the residuals. 
Residual anisotropies can be effectively tested and visualized by mapping the residuals as a function of sky direction.
This can be performed in different ways, from which we present two:
one that can be applied as a post-analysis after determining the dipole parameters by other means (\emph{fixed direction analysis}), and another that allows for an independent estimation of the dipole (\emph{full-sky analysis}). We now define the central quantities used in this method. 

Given a set of residuals and a test direction $\hat{n}$, one can define the Pearson correlation coefficient between the residuals and $\hat{n}$ as:
\begin{equation}
     r(\hat{n}) \equiv \text{corr}(\text{res}_i,\;\cos\theta_i(\hat n)) = \frac{\text{cov}(\text{res}_i,\;\cos\theta_i(\hat n))}{\sigma(\text{res}_i)\, \sigma(\cos\theta_i(\hat n))}
     \label{eq: pearsonr}
\end{equation}
where $\theta_i(\hat{n})$ is the separation angle between $\hat n$ and the direction $\hat{n}_i$ of the $i$th GRB. Also, $\text{corr}$, $\text{cov}$ and $\sigma$ stands for the correlation, covariance and standard deviation. 

By assuming a pure dipolar form for the residuals as:
\begin{equation}
    \text{res}_i = A_d \cos(\vartheta_i) +\epsilon = A_d \;(\hat{d}\cdot \hat{n}_i) +\epsilon\;,
    \label{eq: res dip}
\end{equation}
where $\epsilon$ stands for a random noise with zero mean and uncorrelated with sky directions, $\hat{d}$ represents the dipole direction, $A_d$ the dipole amplitude and $\vartheta_i$ the angle between the dipole $\hat{d}$ and $\hat{n}_i$ (as defined in \eqref{eq:vartheta2}), equation \eqref{eq: pearsonr} becomes:
\begin{align}
    r(\hat n) 
    &= \frac{\text{cov}(A_d \;(\hat{d}\cdot \hat{n}_i), (\hat{n}_i\cdot\hat{n}))}{\sigma(A_d \;(\hat{d}\cdot \hat{n}_i))\, \sigma(\hat{n}_i\cdot\hat{n})} \label{eq: r dipres}\\
    &= \text{sign}(A_d) \; \text{corr}((\hat{d}\cdot \hat{n}_i), (\hat{n}_i\cdot\hat{n}))\;,
    \label{eq: r A cos1}
\end{align}
where $\text{sign}(A_d) = \pm$ depending on the sign of $A_d$. In the second equality, we have made use of the identities $\text{cov}(AX,Y) = A\;\text{cov}(X,Y)$ and $\sigma(AX) = |A|\;\sigma(X)$. 

In case the source distribution is sufficiently isotropic~\footnote{See Appendix \ref{A: isotropy} for full derivation, discussion and results for the isotropy test of the Combined Sample.}, the Pearson correlation can be simplified to
\begin{align}  
r(\hat n)   = \text{sign}(A_d) \;(\hat{d}\cdot \hat{n}) = \text{sign}(A_d)\;\cos(\angle(\hat{d}, \hat{n}))\;.
\label{eq: r A cos2}
\end{align}
As shown in \eqref{eq: r A cos2}, the correlation vanishes when $\hat{n}$ is orthogonal to $\hat{d}$, forming a \emph{circle of zero correlation} on the sky, perpendicular to the dipole axis. Conversely, the correlation reaches its maximum $(r_{max})$ when $\hat{n}$ aligns with the dipole. This angular behavior enables an independent determination of the dipole direction from the residual pattern alone.

Assuming the test direction to be the dipole direction, $\hat{\mathbf{n}}=\hat{d}$, equation \eqref{eq: r dipres} becomes:
\begin{align}
    r(\hat d) &
    = A_d\;\frac{\text{cov}((\hat{d}\cdot \hat{n}_i), (\hat{n}_i\cdot\hat{d}))}{\sigma(\text{res})\, \sigma(\cos\vartheta_i)} \\
    &= A_d \;\frac{\text{var}(\cos\vartheta_i)}{\sigma(\text{res}_i)\, \sigma(\cos\vartheta_i)} 
     = A_d \;\frac{\sigma(\cos\vartheta_i)}{\sigma(\text{res}_i)} \;,
\end{align}
where we have used the definition $\text{var}(X) = \sigma^2(X)$. Isolating $A_d$, this gives us the following dipole amplitude estimation:
\begin{equation}
    A_d = r(\hat{d}) \cdot \frac{\sigma(\text{res}_i)}{\sigma(\cos\vartheta_i)}.
    \label{eq: A res}
\end{equation}

\paragraph{Fixed direction analysis}
A straightforward way to probe residual anisotropies is to fix a specific preferred direction and examine the residuals as a function of the angular separation $\theta_i$ between each GRB and that direction.
When used as a post-analysis tool--as done in this work--a natural choice for the fixed direction is the best-fit dipole direction $\hat{d}$ obtained from a separate dipole measurement method (here, the dipole fitting method results). 

The residuals are defined as:
\begin{align}
    \text{res}_{\Lambda CDM} = \mu_{grb} - \mu_{\Lambda CDM}\;,
    \label{eq:residualsLCDM}
\end{align}
where $\mu_{grb}$ is given by \eqref{eq:mu grb 2D} with the best-fit $(a,c)$ parameters shown in Figure \ref{fig:CornerPlot2D}.
If \eqref{eq:residualsLCDM} presents a pure dipole, as given in equation \eqref{eq: res dip}, then a linear correlation between $\text{res}_{\Lambda CDM}$ and $\cos\vartheta_i$ must exist. A Pearson test can then be performed to quantify the degree of linear correlation $r$ and its corresponding $p-$value.

\begin{table*}[!ht]
\caption{Table with the best-fit values of the Long, XRR-XRF and the full combined samples, using the dipole fit method and correcting for evolutionary effects. All dipole directions (RA, DEC) are in Equatorial Coordinates.}
\label{Table:Dipole results}
\vspace{0.2cm}
\makebox[\textwidth][c]{
\begin{tabular}{lcccccc}
\multicolumn{7}{c}{\textbf{(a) Dipole Fit Method - MCMC results}} \\
\multicolumn{7}{c}{$\;$}\\
 & \multicolumn{3}{c}{GRB parameters} & \multicolumn{3}{c}{Dipole parameters}\\
 \hline
Sample & $a$ & $c$ & $\sigma_{\rm int}$ & $A_d$ & RA [deg] & DEC [deg] \\
\hline
Long      & $-1.15 \pm 0.11$ & $50.94 \pm 0.44$ & $0.285^{+0.027}_{-0.034}$ & $0.81 \pm 0.38$ & $133^{+20}_{-30}$ & $-17^{+24}_{-29}$ \\
XRR-XRF   & $-0.899 \pm 0.099$ & $49.91 \pm 0.39$ & $0.204^{+0.023}_{-0.028}$ & $0.48^{+0.20}_{-0.38}$ & $157^{+50}_{-80}$ & $-46^{+13}_{-38}$ \\
Combined     & $-1.046 \pm 0.073$ & $50.56 \pm 0.28$ & $0.251^{+0.017}_{-0.020}$ & $0.63^{+0.25}_{-0.23}$ & $134 \pm 30$ & $-36 \pm 21$ \\
\hline
\end{tabular}
}
\makebox[\textwidth][l]{
\begin{minipage}{0.47\textwidth}
\vspace{0.4cm}
\begin{tabular}{lcccc}
\multicolumn{5}{c}{\textbf{(b) Anisotropic Residual - Fixed direction}} \\
\multicolumn{5}{c}{$\;$}\\
 \hline
Sample &  $r_{\Lambda CDM}$ & $p_{\Lambda CDM}$ & $r_{dip}$ & $p_{dip}$ \\
\hline
Long   & 0.354 & 0.001 & 0.067 & 0.547 \\
XRR-XRF   & 0.242 & 0.034 & 0.068 & 0.558 \\
Combined  & 0.282 & $1.5\times10^{-4}$ & 0.054 & 0.475\\
\hline
\end{tabular}
\end{minipage}
\hspace{1.2cm}
\begin{minipage}{0.5\textwidth}
\vspace{0.4cm}
\begin{tabular}{lccc}
\multicolumn{4}{c}{\textbf{(c) Anisotropic Residual - Full sky}} \\
\multicolumn{4}{c}{$\;$}\\
 \hline
Sample &  $A_d$ & RA [deg] & DEC [deg] \\
\hline
Long      & $0.89^{+0.75}_{-0.28}$ & $131\pm 28$ & $-20\pm 25$ \\
XRR-XRF   & $0.56^{+0.52}_{-0.11}$ & $144\pm 32$ & $-54\pm 13$ \\
Combined     & $0.67^{+0.42}_{-0.26}$ & $134 \pm 39$ & $-36 \pm 26$ \\
\hline
\end{tabular}
\end{minipage}
}
\end{table*}

This test can be easily extended to other background cosmological models, including anisotropic ones such as Bianchi models, or inhomogeneous yet isotropic models like Lemaître-Tolman-Bondi (LTB), to evaluate whether they more appropriately capture the features in the data.

Furthermore, in order to compare two different theories (e.g. $\mu_{\text{dip}}$ vs. $\mu_{\Lambda CDM}$), one can repeat the analysis using the ``competing'' background cosmology when defining the residuals. In our case, we have:
\begin{align}
\text{res}_{dip} = \mu_{grb} - \mu_{\text{dip}}.
\end{align}
By comparing the Pearson $r$ and $p$-values obtained using $\mu_{\text{dip}}$ to those from the standard $\Lambda$CDM case, one can assess whether the dipole model provides a better fit. A good fit should leave no statistically significant residual linear correlation.

\paragraph{Full-Sky Anisotropy Mapping}
This method offers an alternative way to independently measure the dipole’s direction and amplitude. 
Here, the dipole direction is no longer fixed. 
Instead, we scan the entire sky, performing the residual analysis at each location.
At each point, the weighted\footnote{We have opted for the weighted correlation coefficient in order to take into account the error bars in each data point. The weight is defined as $1/\sigma^2_{res} = 1/\sigma^2_{\mu_{grb}}$. Note that in this way we guarantee that the intrinsic scatter of the sample is also directly taken into account when performing the anisotropic residual analysis. That said, the results were not much affected by taking this factor into account, due to the error variability being small in this data set.} Pearson correlation coefficient $r$ \citep{Han22} between $\text{res}_i^{\Lambda CDM}$ and $\cos\vartheta_i$ is calculated, along with its corresponding $p-$value. 
This process generates a full-sky map of correlation strength, revealing regions with the strongest alignment and statistically significant correlations.

The dipole direction and amplitude are defined as follows: we first identify the region (or pixels) where the Pearson correlation $r$ is maximal (or minimal, for the antipodal point) and the corresponding $p$-value falls below a chosen significance threshold -- for example, $p=0.05$ ($5\%$ confidence level). 
The dipole direction is obtained taking the weighted average inside this region, with weight given by the p-value of each pixel. 
The same procedure is applied to obtain the maximal correlation $r_{\text{max}}$ and the dipole amplitude $A_d$, using equation \eqref{eq: A res}.

To obtain robust error estimates, we performed 10,000 bootstrap resamplings of the data~\citep{Davison_Hinkley_1997}, applied the same procedure as above to compute the correlation coefficients and dipole amplitudes, and used the $95\%$ confidence interval of the resulting distribution in order to obtain the error bars.

\section{Results}
\label{S: results}

\subsection{Dipole Fitting Results}

\begin{figure*}
\centering
\subfigure[Triangle plot showing the marginalized one-dimensional and two-dimensional posterior distributions for the dipole ($A_d$, RA, DEC) and the GRB standardization (a, c, $\sigma_{int}$) parameters of the Long, XRR-XRF, and combined GRB samples.]{
  \includegraphics[width=0.85\textwidth]{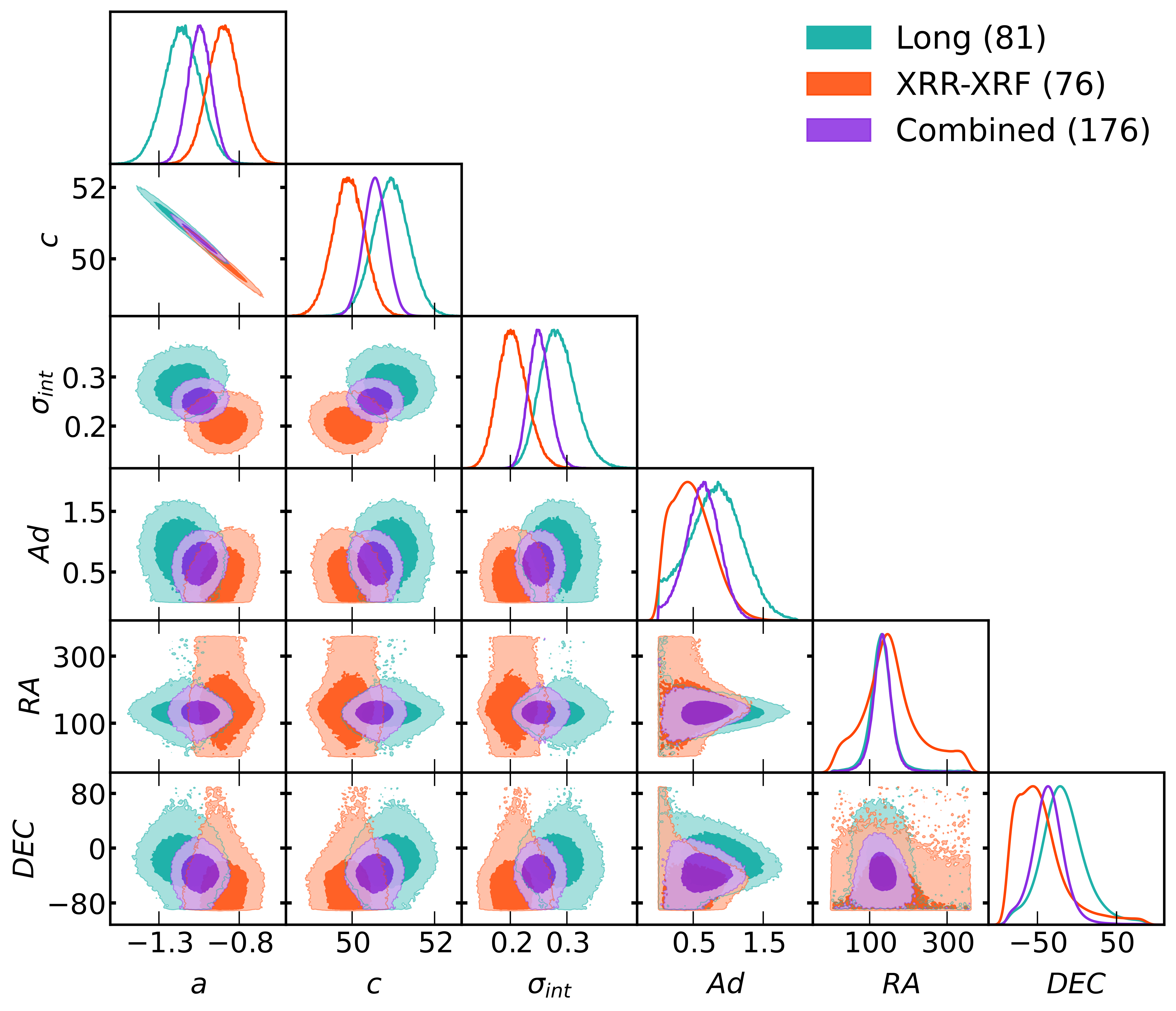}
  \label{fig:dipole_multi2D}
}
\hfill
\subfigure[Directions of reported dipole anisotropies in the sky in Galactic ($\ell$, b) coordinates on a Mollweide projection. Different markers correspond to measurements from the literature, with shaded ellipses indicating their quoted $1\sigma$ uncertainties, with text labels. The result of this analysis is shown in orange, while the (magenta) star indicates the direction of the CMB dipole.]{
  \includegraphics[width=0.85\textwidth]{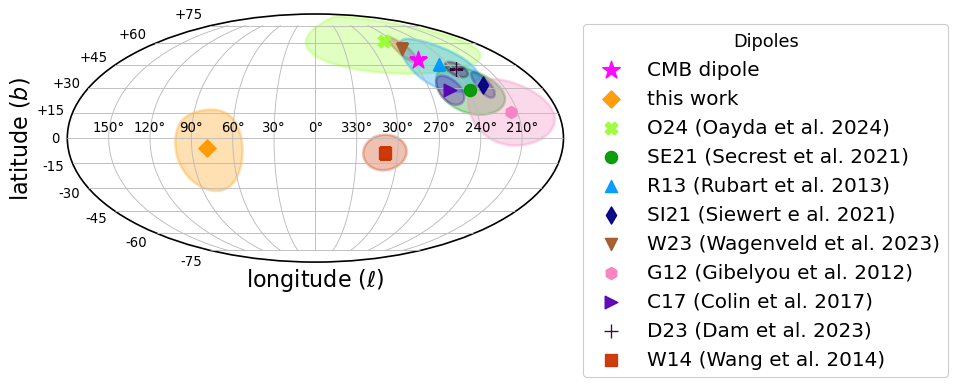}
  \label{fig:sky_dipoles}
}
\label{fig:multimol}
\caption{Bayesian analysis results and Mollweide map showing the obtained direction together with other anisotropy studies. }
\end{figure*}

We apply the dipole model \eqref{eq: mu dipole} to the GRB datasets and show the results for the Long (81 GRBs), XRR-XRF (76 GRBs) and full Combined sample (176). The analysis is performed within a Bayesian statistical framework. We adopt flat priors for all model parameters and employ the Affine-Invariant Markov Chain Monte Carlo (MCMC) ensemble sampler introduced by \citet{Foreman-Mackey_2013} to constrain the posterior distributions. The direction of the dipole in Equatorial coordinates was chosen to give a uniform distribution over the entire sphere, for $0^{\circ} < RA < 360^{\circ}$ and $-90^{\circ} < DEC < 90^{\circ}$. For the dipole amplitude, we choose uniform priors over the range $A_d\in [0, 2]$. In Figure \ref{fig:dipole_multi2D}, we show the marginalized posterior distributions and parameter correlations, where the shaded areas denote the $1\sigma$ and $2\sigma$ errors propagated from the uncertainties of the GRB distance moduli as defined in Equation \eqref{eq:sigma_mu}. These results can also be found in Table \ref{Table:Dipole results}(a).
The GRB's luminosity distances are standardized by the L-T relation method~\eqref{eq:mu grb 2D}, with free parameters given by $a, c$ and the intrinsic scatter $\sigma_{int}$. 
We further account for redshift evolution following the prescription outlined in Equation \eqref{eq:evolution}. In order to verify the dependency of our results on selection effects due to flux thresholds, we have performed the dipole fit with and without taking the redshift evolution into account. The results are consistent within $1\sigma$ for all samples, and the posterior distributions are better constrained when taking the evolution into account. The analysis simultaneously constrains the GRB empirical relation parameters and the dipole parameters ($A_d$, RA, DEC).
The different colors in Figure \ref{fig:dipole_multi2D} denote the different GRB datasets used in this analysis.
We have not included the results for Platinum alone because, as discussed in Appendix \ref{A: Platinum sample}, its source distribution is not optimal for constraining the dipole. Our results converge within $1\sigma$ between the different sub-samples for the dipole parameters. However, this is not the case for the GRB parameters, where it was found up to $2\sigma$ difference for the standardization of the $a, c, \sigma_{int}$ values between the Long and the XRR-XRF subclasses. This fact denotes the different characteristics among the GRB classes and the importance of fitting the GRB standardizing parameters together with the dipole. 
Furthermore, one can notice the high constraining power of the intrinsic scatter in the case of the full Combined sample, denoted in purple, compared to the other two samples. In order to test for effects of local structures in the obtained dipole signal, we have performed the same analysis imposing a low redshift cut of $z_{min} = 0.1$. The obtained results were all within $1\sigma$ with the ones here presented, so we have not included them in the paper.

Figure \ref{fig:sky_dipoles} shows the on-sky 
velocity direction of our analysis (orange marker), in galactic coordinates, on a Mollweide projection.
As described in Appendix \ref{A: boost}, by using the dipole model \eqref{eq: mu dipole}, the boost velocity between the observer and the GRB rest-frame lies in the direction antipodal to the dipole direction.
For comparison, we also display recent high-redshift dipole measurements from the literature using a variety of tracers and statistical techniques (see reviews ~\cite{2023CQGra..40i4001A, 2022AnPhy.44769159P}). The magenta star denotes the CMB dipole direction, while other symbols show reported dipole directions and their quoted $1\sigma$ uncertainties. Several studies have reported large-scale dipole anisotropies using different datasets: quasars from the CatWISE sample (~\cite{Secrest:2020has}, SE21) and a Bayesian reanalysis (~\cite{Dam:2022wwh}, D23); cosmic radio sources from the NVSS survey (~\cite{rubart2013, 2012MNRAS.427.1994G}, R13, G12) and the combined TGSS/NVSS radio catalogues (~\cite{Colin:2017juj, Siewert:2020krp}, C17, SI21). Additional analyses have employed SNeIa and GRBs (~\cite{Wang:2014vqa}, W14), while more recent radio-based measurements (\cite{Wagenveld:2023kvi, Oayda:2024hnu}, W23, O24), yield directions close to the CMB dipole. Overall, while most directions cluster broadly around the CMB dipole, the reported amplitudes vary significantly across datasets and methods. Possible reasons for the discrepancies with the results of the present analysis are discussed in the Section \ref{S: conclusion}.

The majority of high-redshift studies have used the Ellis \& Baldwin test (number count) to determine the dipole anisotropy, exceeding the expectations from the standard model at about $2-3 \sigma$ level \citep{Colin:2017juj, Secrest:2020has, singal2019}. Our results agree with these studies in the sense of also finding a $2-3 \sigma$ discrepancy. However, a direct comparison of our dipole amplitude with their results is highly non-trivial due to the difficulty arising in interpreting the amplitude in the dipole fitting method as a boost velocity (see \ref{A: boost} for a full discussion). Moreover, when comparing to the GRB dipole results obtained from the analysis of \citep{10.1093/mnras/stac498}, their dipole amplitude is one order of magnitude lower than ours. The reasons for that could be due to the different sample, GRB standardization method and one extra degree of freedom in their fit parameters.

\subsection{Anisotropic Residual Results}
\label{S: residual analysis}

\begin{figure}
\subfigure[Fixed Direction Residual Analysis]{
\centering
  \includegraphics[width=0.43\textwidth]{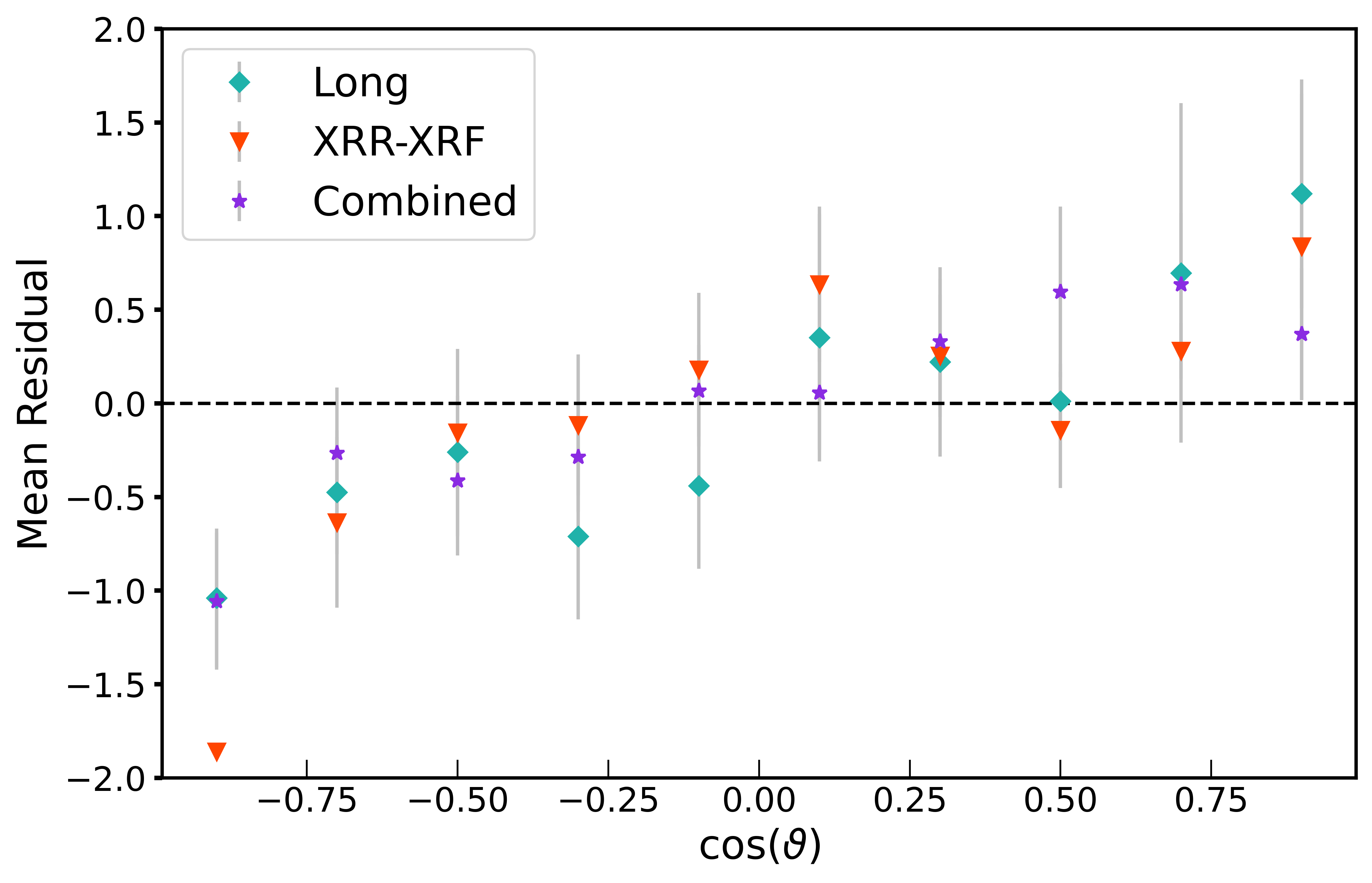}
  \label{fig:anisotropic_pearson}
}
\hfill
\subfigure[Full-Sky Anisotropy Maps]{
  \includegraphics[width=0.43\textwidth]{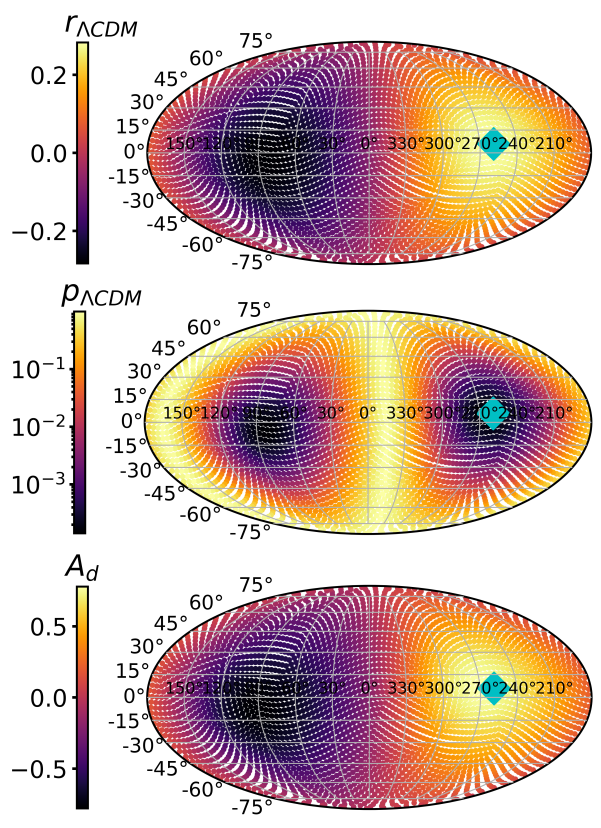}
  \label{fig:r-mollweide}
  \hfill
}
\hfill
\subfigure[Likelihood of maximal dipole amplitude for the 10,000 isotropic simulations.]{
\centering
  \includegraphics[width=0.4\textwidth]{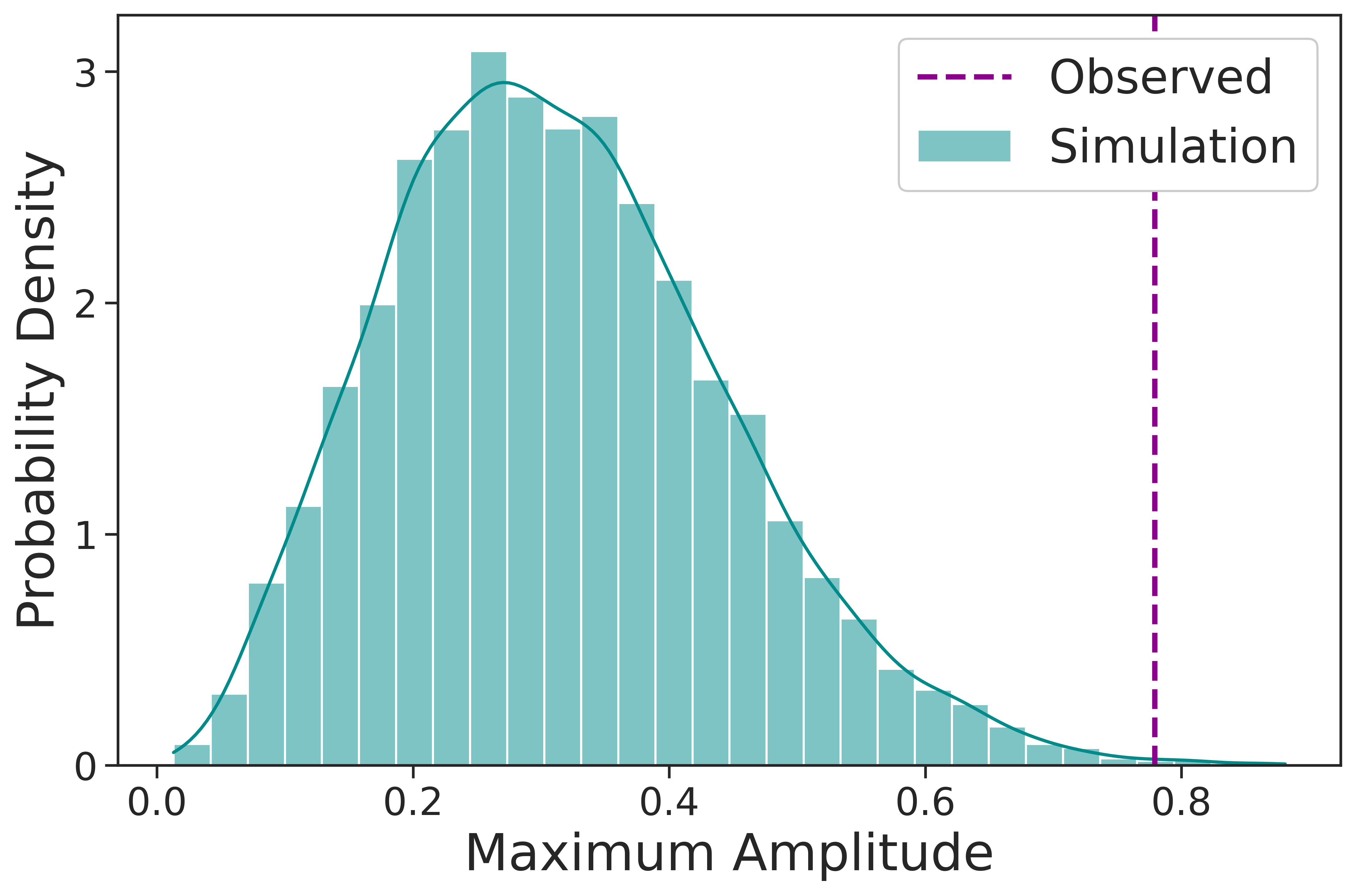}
  \label{fig:simulation}
}
\caption{Anisotropic residual results}
\label{fig: anisotropic residuals}
\end{figure}
We now present the results of the anisotropy tests described in Section \ref{S: Residual Analysis theory}.

\paragraph{Fixed Direction Analysis}
We have applied the fixed direction test as a post-analysis to the Dipole Fit results obtained using MCMC. To do so, first we define the residuals $\text{res}_{\Lambda CDM}$, as in \eqref{eq:residualsLCDM}, with respect to $\mu_{\Lambda CDM}$. 
The angular separation $\cos\vartheta_i$ (equation \eqref{eq:vartheta2}) was defined using the best-fit right ascension (RA), and declination (DEC) obtained from the Bayesian analysis (see Table \ref{Table:Dipole results}(a)).
To compute the degree of linear correlation, we performed a Pearson correlation test between $\text{res}_{\Lambda CDM}$ and $\cos\vartheta_i$.
Afterwards, we have performed the same test, keeping $\cos\vartheta_i$ the same, but now calculating its linear correlation with $\text{res}_{dip}$, given by \eqref{eq: res dip}, where $\mu_{dip}$ is calculated using the best-fit dipole parameters from Table \ref{Table:Dipole results}(a).

Figure~\ref{fig:anisotropic_pearson} shows the linear trend between $\text{res}_{\Lambda CDM}$ and $\cos\vartheta_i$ for the Long, XRR-XRF and Combined samples. For improved visualization, we have divided $\cos\vartheta_i$ into 10 bins, each point in the figure representing the mean residual in each bin. The linear trend is clearly visible for all presented samples. 

The resulting Pearson $r$ and $p$-values for both $\text{res}_{\Lambda \text{CDM}}$ and $\text{res}_\text{dip}$ are reported in Table~\ref{Table:Dipole results}(b). The results show that, besides the obtained Pearson correlation coefficients $r_{\Lambda CDM}$ not supporting an obvious linear correlation, the $p-$values obtained, specially for the Combined sample, indicate the opposite.
This suggests that some directional dependence remains unaccounted for in the standard $\Lambda$CDM model. When incorporating the dipole correction in $\text{res}_{\text{dip}}$, the correlation effectively vanishes, with both $r$ and $p$-values pointing to a lack of significant anisotropy. This supports the conclusion that the dipole model provides a better description of the data by removing the directional residual signal.

\paragraph{Full-Sky Anisotropy Mapping}
To further explore directional dependence without assuming a fixed dipole orientation, we applied the full-sky scanning method described in Section~\ref{S: dipole}. Figure~\ref{fig:r-mollweide} displays the resulting Mollweide projections of the Pearson $r_{\Lambda CDM}$-values, its corresponding $p_{\Lambda CDM}$-values and the obtained dipole magnitude $A_d$ across the sky.

The maps reveal a clear dipolar pattern, with a well-defined region of maximal correlation. The best-fit direction derived from this scan aligns well with the direction obtained from the Bayesian dipole fit (indicated by the turquoise marker), demonstrating consistency between the two independent methods. 
We present the results for the Long, XRR-XRF and Combined samples in Table \ref{Table:Dipole results}(c). The matching of these results with the Bayesian analysis presented in Table \ref{Table:Dipole results}(a) strengthens the case for a dipolar anisotropy in the GRB data and validates the effectiveness of the proposed \emph{Anisotropic Residual Analysis Method} as an independent approach for detecting large-scale directional features.

To further assess the results, we ran 10,000 simulations using the Combined sample, preserving GRB properties (flux, redshift, luminosity) while randomizing sky positions.
We applied the same pipeline described above to compute the dipole amplitude sky map for each run. Since in many simulations the p-value never fell below $p=0.05$ ($5\%$ confidence level) due to lack of dipole signal, instead of calculating the dipole amplitude as in Section \ref{S: Residual Analysis theory}, we recorded the maximum obtained amplitude. Figure \ref{fig: anisotropic residuals}(c) shows the probability distribution of these maxima across the 10,000 runs. The dashed magenta line marks the maximum value obtained with the Combined sample $(A_{\max}=0.779)$, whose p-value from the simulations was $p=0.0016$.

We then ran additional 10,000 simulations, this time reassigning GRB positions while preserving the original angular distribution. This was achieved by computing the data power spectrum and corresponding density map, and assigning pixel probabilities accordingly (see Appendix \ref{A: isotropy}). The results remained essentially unchanged, yielding a new p-value of $p=0.0011$. These simulations confirm the robustness of the signal: whatever its origin, it cannot be attributed to the sky distribution of the sample.

\section{Discussion and Conclusion}
\label{S: conclusion}

The cosmic dipole tension is currently one of the most controversial topics in observational cosmology. Recent analyses (e.g. \cite{Dam:2022wwh}) report a discrepancy exceeding $3\sigma$ challenging the cosmological principle underlying the standard $\Lambda$CDM model. High-redshift probes (quasars, GRBs, radio galaxies) generally yield dipole directions consistent with the CMB dipole (see Figure~\ref{fig:sky_dipoles}), but with amplitudes that exceed the kinematic expectation ~\citep{2023CQGra..40i4001A}. The tension remains, although the latest GRB analysis finds no significant dipole \citep{Luongo:2025}.

We present the first measurement of the cosmic dipole using \emph{Swift} data and the first using the well established L-T relation of GRBs (see Section \ref{S: standardizing}). Our analysis employs 176 long GRBs observed by Swift, standardized with the L-T (Dainotti) relation and corrected for redshift evolution, representing nearly $50\%$ more sources than previous dipole studies based on the Amati relation \citep{10.1093/mnras/stac498}. As presented in Appendix \ref{A: isotropy}, our source angular distribution is isotropically distributed in the sky. Using both the Dipole Fit Method and a newly introduced Anisotropic Residual Analysis, we find consistent evidence for a dipolar modulation in the GRB distance modulus. Our extensive simulations also confirm that this signal cannot be attributed to chance alignments or anisotropic sampling. Furthermore, incorporating the dipole term eliminates residual correlations, showing that it provides a better description of the data than isotropic $\Lambda$CDM.

While the dipole direction is aligned approximately towards CMB dipole direction, the boost velocity of the observer reference frame connected with such a dipole is actually in the antipodal direction (see Appendix \ref{A: boost} and Figure \ref{fig:sky_dipoles}). The direction found, therefore, is not consistent with earlier studies. The reason for such a discrepancy is yet not well understood.
First of all, this could be due to a non-isotropic exposure map across the sky. 
The \emph{Swift} telescope is composed of the BAT (Burst Alert Telescope), XRT (X-Ray Telescope) and UVOT (Ultraviolet/Optical Telescope) telescopes. BAT is known to present a non-isotropic sky coverage \citep{Lien:2025zzo} due to off-axis sensitivity, partial coding fraction and anisotropic exposure across the sky.
Furthermore, XRT is a focusing X-ray telescope, with observations almost entirely targeted and not subjected to the same anisotropic exposure features as BAT. For XRT, possible sources of anisotropy come again from anisotropic exposure maps, now due to some bad pixels (due to micrometeoroid hitting a CCD camera), together with some sky regions accumulating deep exposures (fields of repeated visits). While XRT exposure maps do exist, they are produced per observation (per pointing / per ObsID) rather than an all-sky map. 
All these effects can induce Malmquist bias effects: in any direction where the effective flux limit is lower (easier detection), the average GRB will appear brighter, implying a higher detection fraction and leading to a spurious dipole pointing toward high-exposure regions.
A future follow-up to this work is to properly build an all-sky XRT exposure map for the events in our catalog in order to measure their impact on the observed signal. 

Overall, the results here presented establish GRBs as a powerful new probe of large-scale anisotropy at high redshift. They pave the way for joint analyses with low-redshift sources such as SNe Ia, enabling multi-scale tests of the cosmological principle and offering new insights into the origin of cosmic anisotropies. 

\section*{Acknowledgments}
The authors would like to thank Biagio De Simone for all the help with the dataset in the beginning of this work. JS would like to thank Basheer Kalbouneh, Cristian Marinoni, Harry Goodhew and Calvin Chen for useful discussions and inputs.

JS acknowledges the support from the Taiwan National Science and Technology Council, grant No. 112-2811-M-002-132 and Agence Nationale de la Recherche under the grant ANR-24-CE31-6963-01, and the French government under the France 2030 investment plan, as part of the Initiative d’Excellence d'Aix-Marseille Université -  A*MIDEX (AMX-19-IET-012). 
KA acknowledges the support from the Hellenic Foundation for Research and Innovation (H.F.R.I.), under the ``First Call for H.F.R.I. Research Projects to support Faculty members and Researchers and the procurement of high-cost research equipment Grant'' (Project Number: 789). 
MGD acknowledges the support of the DoS and by JSPS Grant-in-Aid for Scientific Research (KAKENHI) (A), Grant Number JP25H00675. 
PC acknowledges the support by Taiwan’s National Science and Technology Council (NSTC) under project number 110-2112-M-002-031, and by the Leung Center for Cosmology and Particle Astrophysics (LeCosPA), National Taiwan University (NTU).

\appendix

\section{The Platinum Sample}
\label{A: Platinum sample}

\begin{figure}
\subfigure[Number of data points per quadrant]{
  \includegraphics[width=0.48\textwidth]{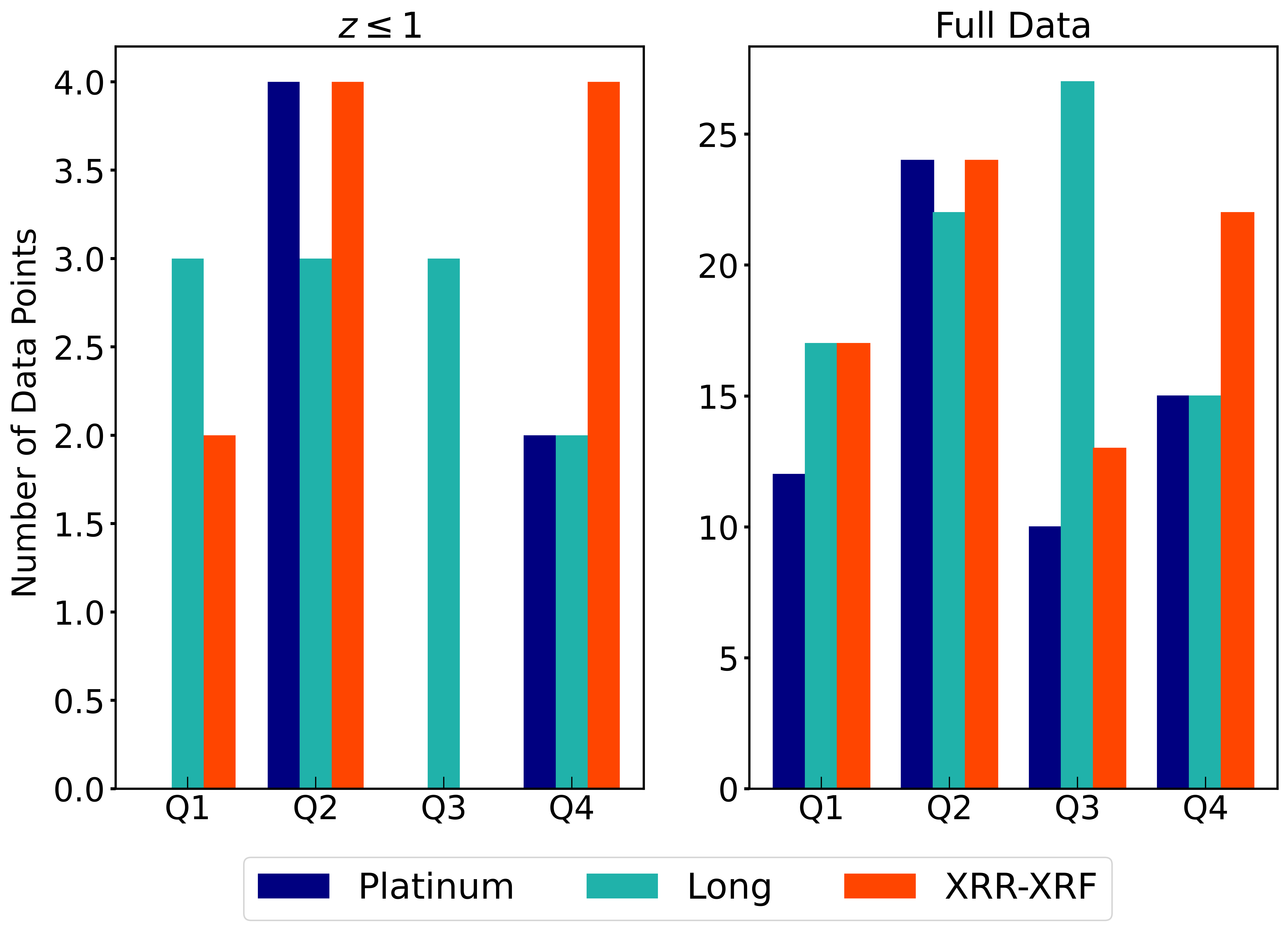}
  \label{fig:QD_z1}
}
\hfill
\subfigure[Mollweide plot for the $z\leq 1$ cut sample]{
  \includegraphics[width=0.48\textwidth]{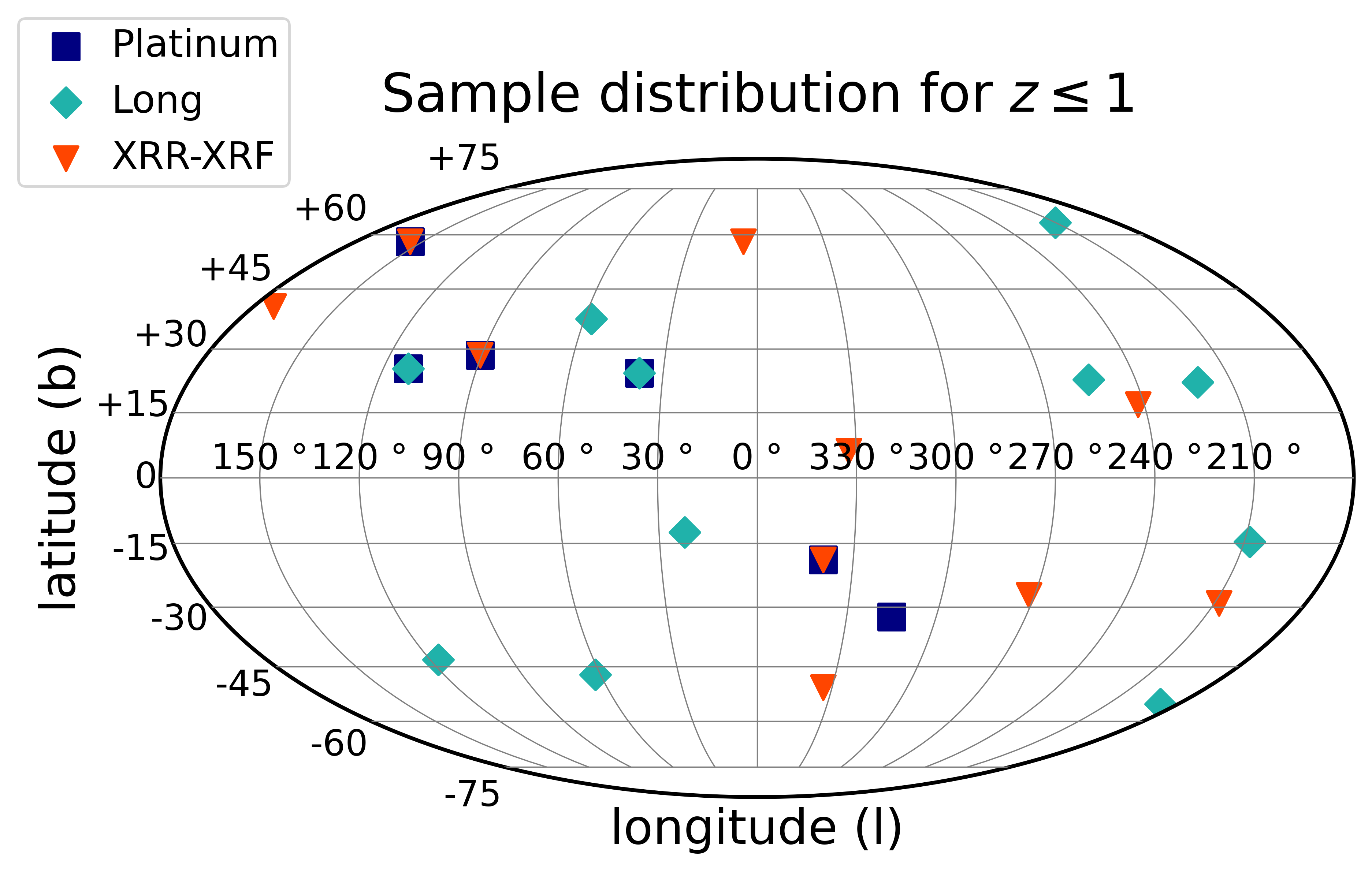}
  \label{fig:samplez1}
}
\caption{Number counts per Quadrant (left) and Mollweide projection (right) for the individual GRB subsamples.}
\label{fig:individual_sample}
\end{figure}

In this Appendix, we discuss the distribution of the Platinum sample and explain why its individual analysis was not included in the main results.

When considering this sample in isolation—without combining it with others—the derived results were found to be inconsistent with those obtained for the Long, XRR–XRF, and Combined samples. This discrepancy motivated a closer examination of the Platinum sample distribution. The analysis revealed a pronounced anisotropy in the source distribution, particularly at low redshifts. To quantify this effect, we counted the number of GRBs in each quadrant of the sky, defined as follows: Q1 spans latitudes $(0, +90^{\circ})$ and longitudes $(180^{\circ}, 360^{\circ})$, Q2 spans $(0, -90^{\circ})$ and $(180^{\circ}, 360^{\circ})$, Q3 spans $(0, +90^{\circ})$ and $(0,180^{\circ})$, and Q4 spans $(0, -90^{\circ})$ and $(0, 180^{\circ})$. The results of this source-counting procedure are shown in Figure \ref{fig:QD_z1} for both $z \leq 1$ and for the full dataset.

Importantly, the dipole direction obtained for the other individual samples aligns with the axis defined by quadrants Q1 and Q3. However, for $z \leq 1$, the Platinum sample contains no events in this region. This can also be seen in the Mollweide projection of Figure \ref{fig:samplez1}, which displays the distribution of all samples at $z \leq 1$. Even when considering the entire redshift range, the number of Platinum GRBs along the dipole axis remains systematically lower than in the other samples (see Figure \ref{fig:QD_z1}).

The impact of missing data along the dipole axis, and its consequences for dipole estimation, has already been investigated in \citet{Oayda:2024voo}. These findings support our interpretation that the anisotropic distribution of the Platinum sample likely explains the inconsistent results obtained when analyzing it separately.

\section{Discussion on the sample isotropy}
\label{A: isotropy}
In Section \ref{S: Residual Analysis theory}, when deriving equation \eqref{eq: r A cos2}, we pointed out the requirement of a sufficiently isotropic sample. Here, we provide a full derivation of this equation, clarify the meaning of isotropy in this context and present a test on the isotropy of the angular distribution for the Combined sample. 

\begin{figure}
  \centering
\includegraphics[width=0.48\textwidth]{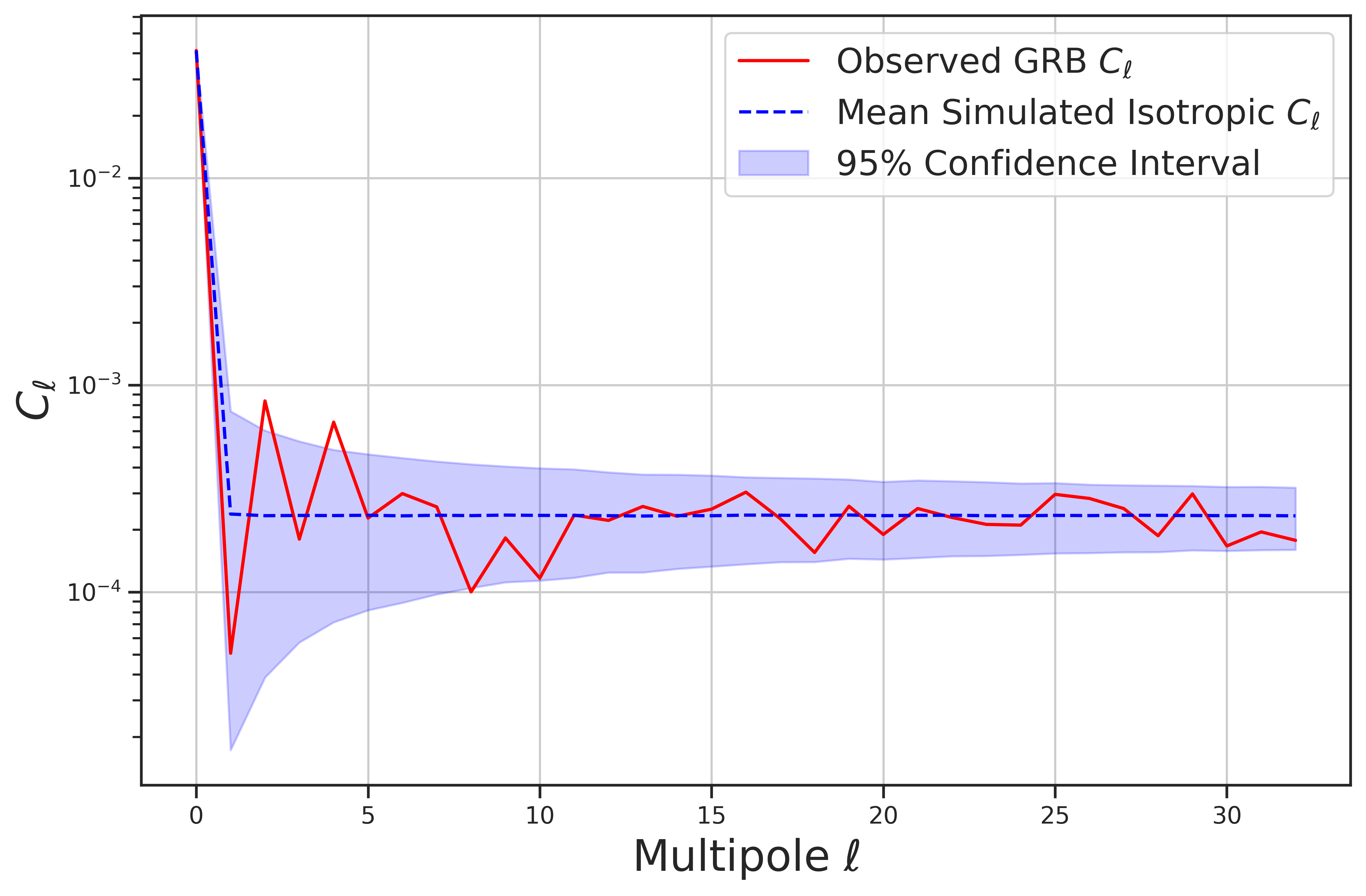}
\caption{Caption}
\label{fig:Comb_sample_isotropy}
\end{figure}

By isotropy, we require that the sky distribution of the data points (vectors $\hat{n}_i$) be such that, for any fixed vector $\vec{v}$, the expectation value $E[(\vec{v}\cdot\hat{n}_i)]$ is approximately zero. Under this assumption, starting from equation \eqref{eq: r A cos1}, we obtain:
\begin{align}
    r(\hat n) 
    &= \frac{\text{cov}(A_d \;(\hat{d}\cdot \hat{n}_i), (\hat{n}_i\cdot\hat{n}))}{\sigma(A_d \;(\hat{d}\cdot \hat{n}_i))\, \sigma(\hat{n}_i\cdot\hat{n})} = \text{sign}(A_d)\;\frac{E[(\hat{d}\cdot \hat{n}_i)(\hat{n}\cdot\hat{n}_i)]}{\sigma(\hat{d}\cdot \hat{n}_i)\, \sigma(\hat{n}_i\cdot\hat{n})}\;,
    \label{eq: r E}
\end{align}
where the covariance definition $\text{cov}[X,Y] = E[XY] - E[X]E[Y]$ has been used. 

It is a well known result that if $\hat{n}_i$ is uniformly distributed over the sphere, given two fixed unit vectors $\hat{a}$ and $\hat{b}$, we have:
\begin{equation}
    E[(\hat{a}\cdot \hat{n}_i)(\hat{b}\cdot \hat{n}_i) ] = \frac{1}{3}(\hat{a}\cdot \hat{b})\;.
    \label{eq:Expec}
\end{equation}
Furthermore, given that the variance in terms of the expectation value is defined as:
\begin{equation}
    \sigma^2[(\hat{a}\cdot \hat{n}_i)] = E[(\hat{a}\cdot \hat{n}_i)^2] - (E[(\hat{a}\cdot \hat{n}_i)])^2\;,    
    \label{eq: sigma2 E}
\end{equation}
under the above mentioned isotropy definition, we have:
\begin{equation}
    \sigma^2[(\hat{a}\cdot \hat{n}_i)] = E[(\hat{a}\cdot \hat{n}_i)^2] = \frac{||a||^2}{3} = \frac{1}{3}\;.   
    \label{eq:sigma2}
\end{equation}
Applying equations \eqref{eq:sigma2} and \eqref{eq:Expec} to \eqref{eq: r E}, we have:
\begin{align}
    r(\hat n) 
    &= \text{sign}(A_d) \frac{(\hat{d}\cdot \hat{n})/3}{\sqrt{1/3} \;\sqrt{1/3}} = \text{sign}(A_d) \;(\hat{d}\cdot \hat{n})\;,
    \label{eq: r Exp}
\end{align}
which recovers equation \eqref{eq: r A cos2}.

The isotropy condition was explicitly tested for the Combined sample by estimating the magnitude of the first-order correction term. This contribution was found to be more than two orders of magnitude smaller than the measured amplitude, thereby confirming the validity of equation \eqref{eq: r A cos2} for our dataset.

An alternative approach to testing isotropy involves analyzing the angular power spectrum of the source distribution and comparing it with that of purely uniform random realizations. To this end, we performed 10,000 simulations of uniformly random sky distributions, computed their respective $C_{\ell}$ spectra, and determined the corresponding $95\%$ confidence interval. The results, presented in Figure \ref{fig:Comb_sample_isotropy}, show that the observed distribution largely falls within this interval, consistent with isotropy.

\begin{figure*}[ht!]
\subfigure{
  \includegraphics[width=0.48\textwidth]{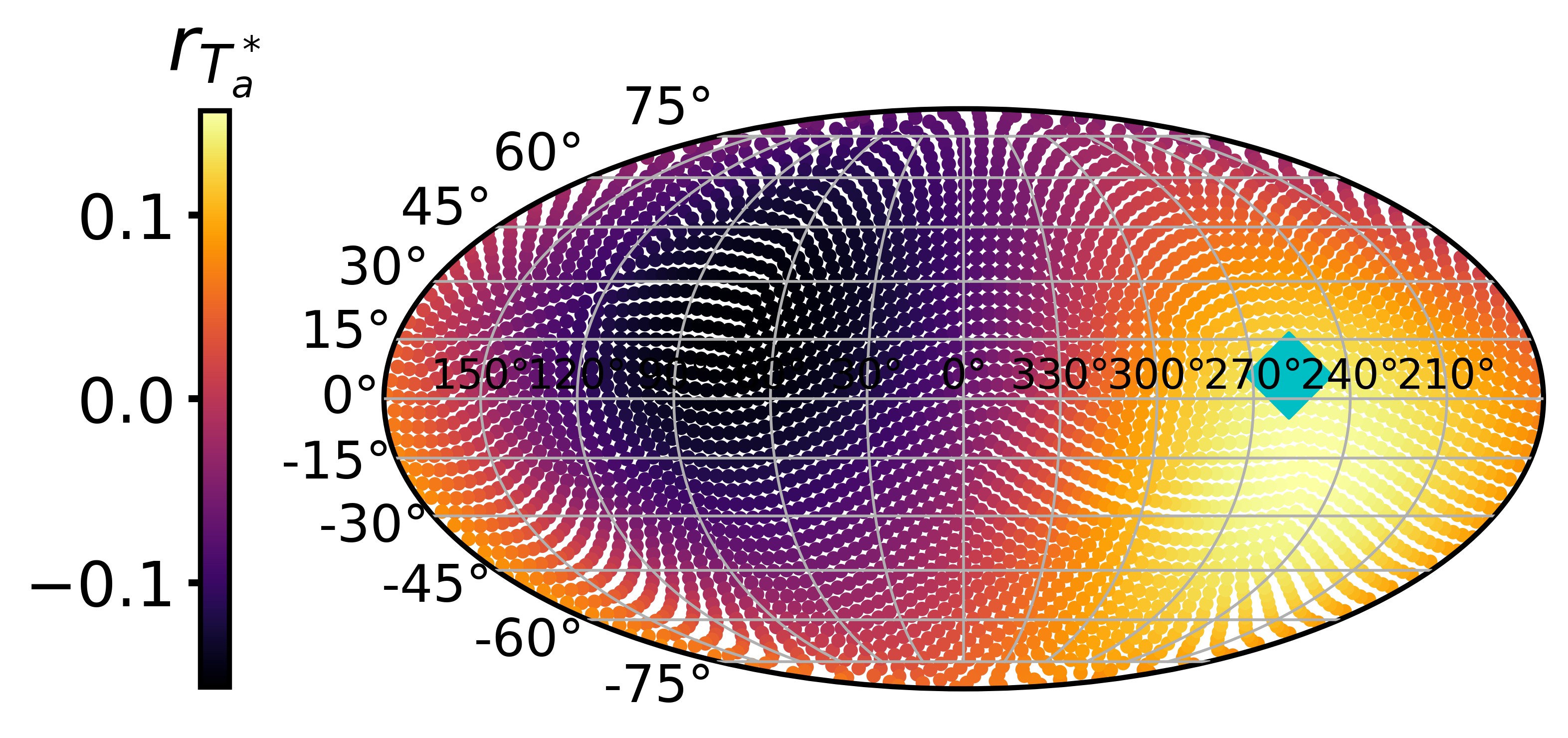}
  \label{fig:Tar}
}
\hfill
\subfigure{
  \includegraphics[width=0.48\textwidth]{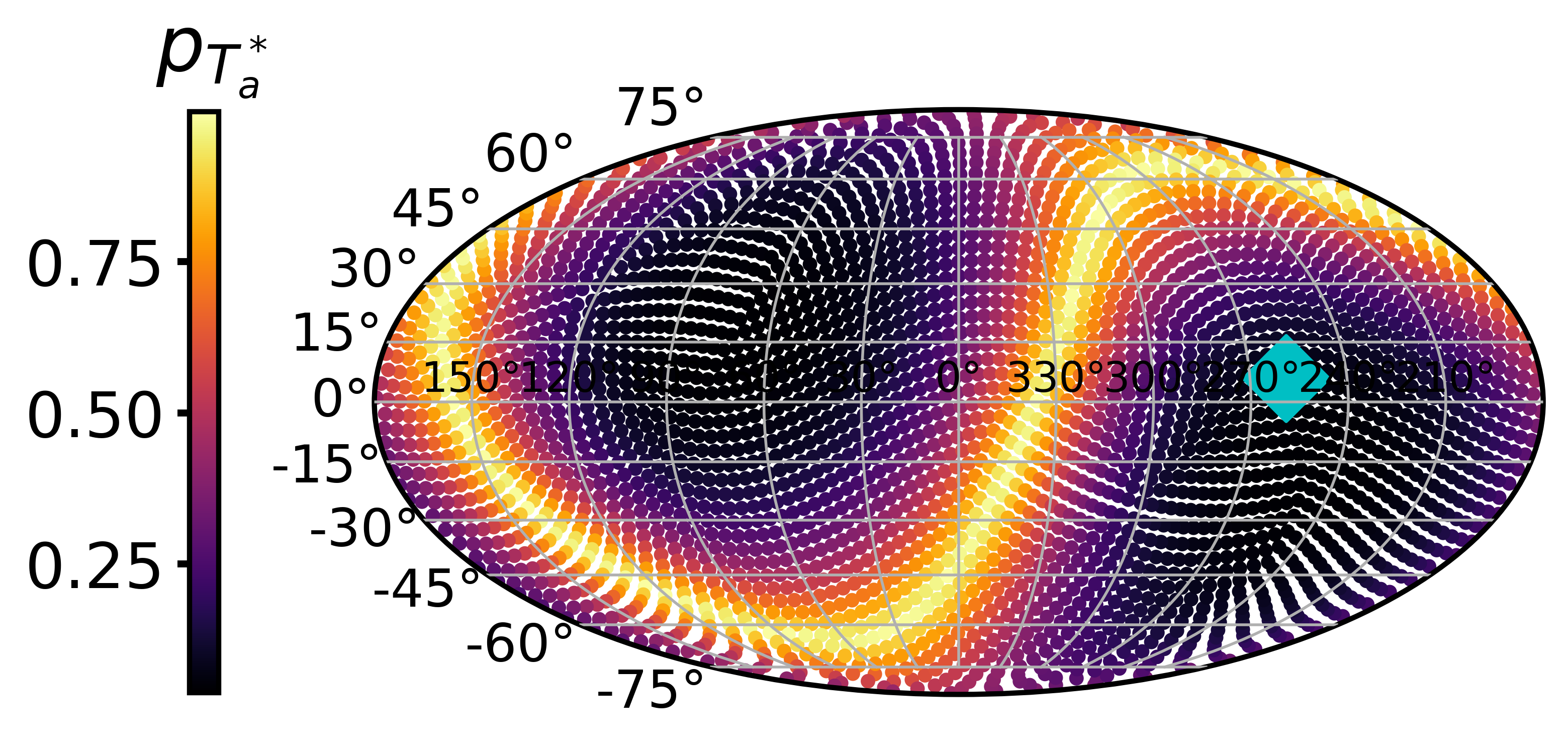}
  \label{fig:Tap}
}
\subfigure{
  \includegraphics[width=0.48\textwidth]{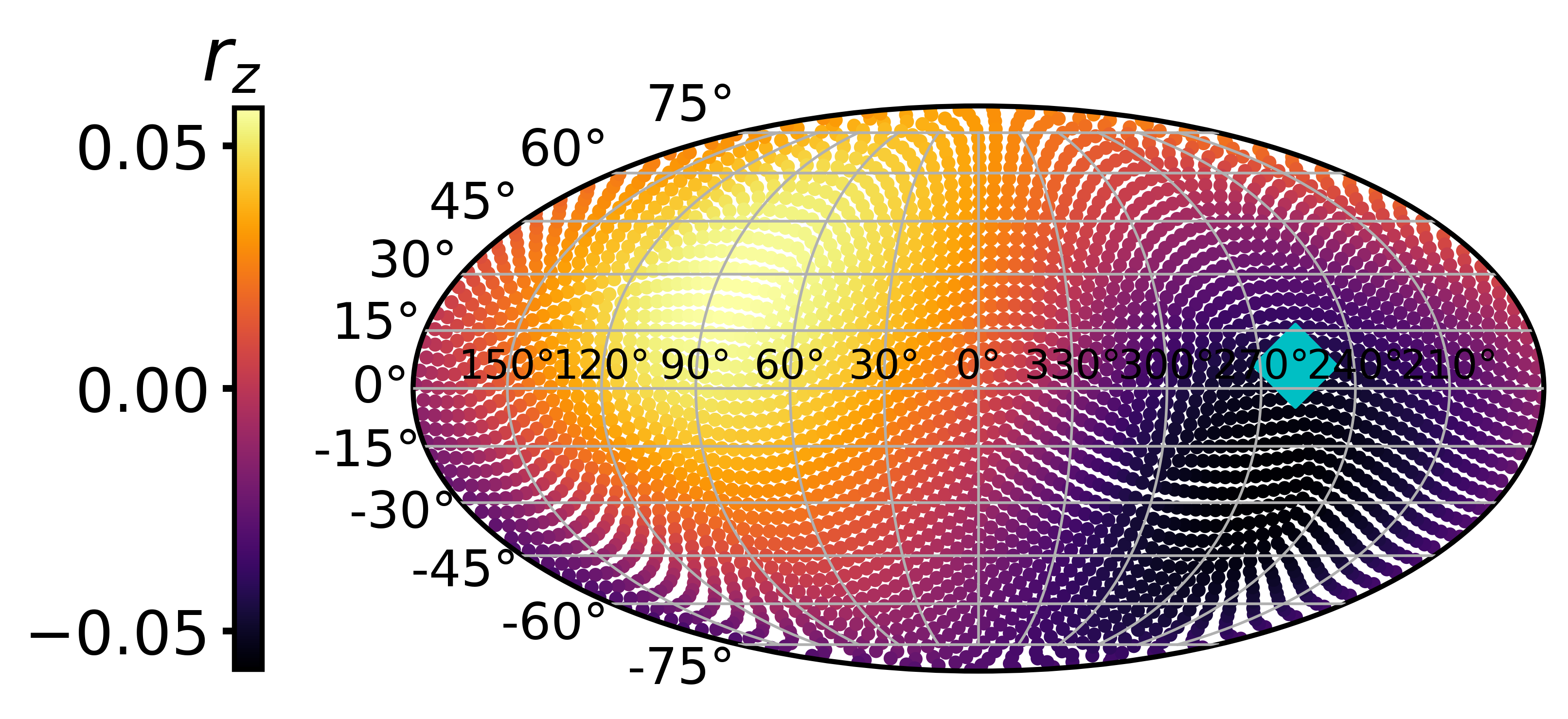}
  \label{fig:zr}
}
\hfill
\subfigure{
  \includegraphics[width=0.48\textwidth]{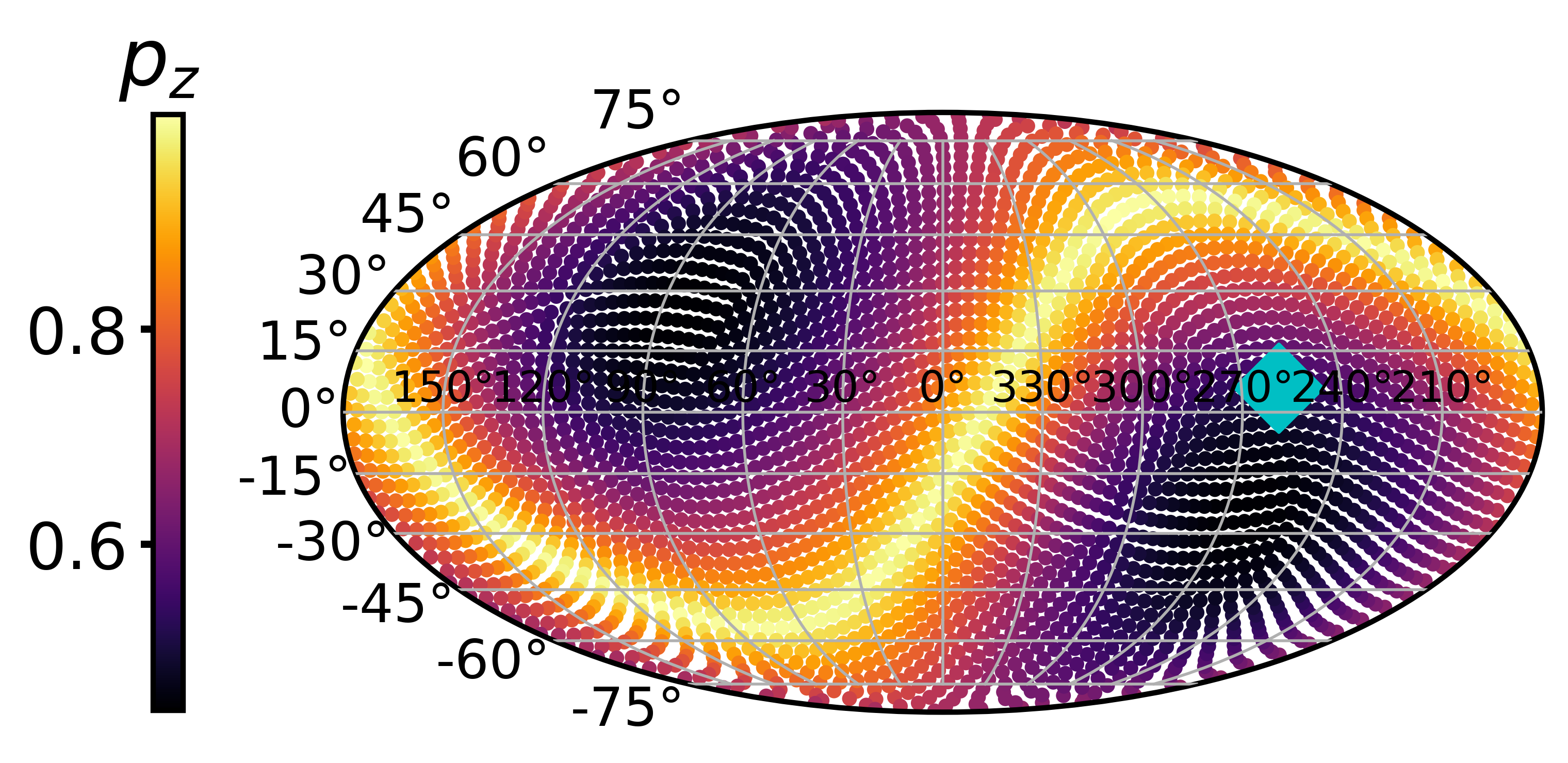}
  \label{fig:zp}
}
\subfigure{
  \includegraphics[width=0.48\textwidth]{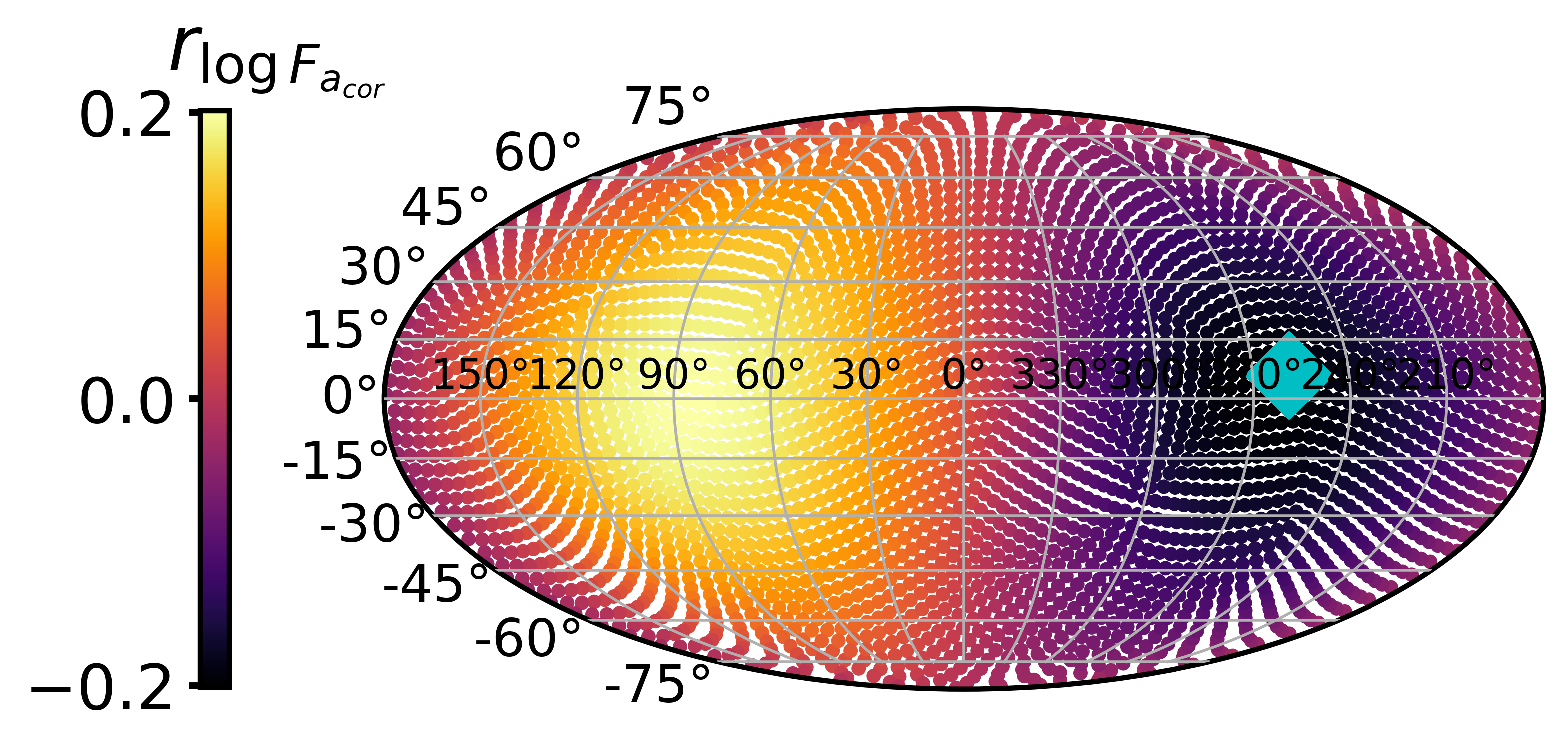}
  \label{fig:Far}
}
\hfill
\subfigure{
  \includegraphics[width=0.48\textwidth]{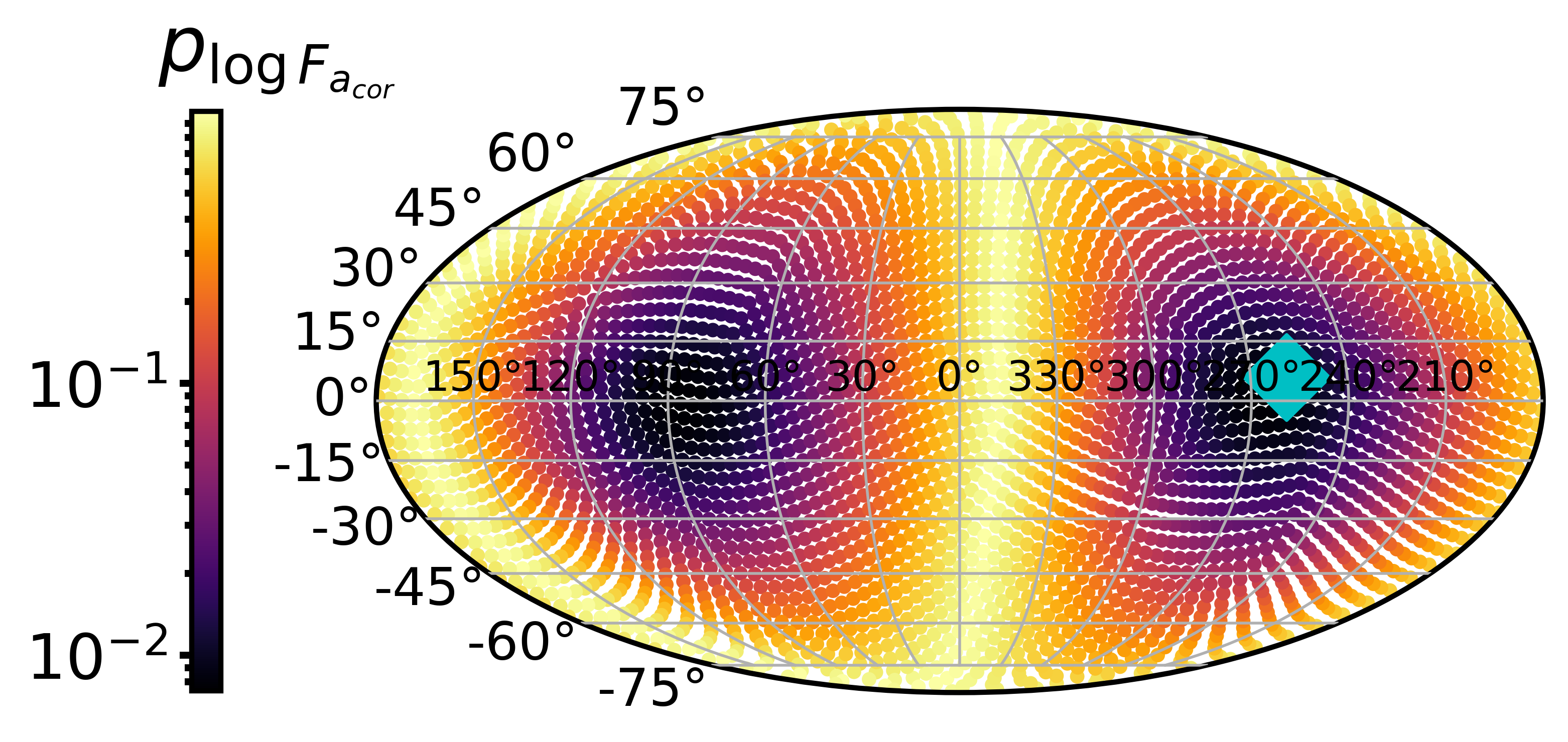}
  \label{fig:Fap}
}
\caption{Full-Sky Anisotropic Residual Analysis applied to radiative properties of the Combined Sample. The left column presents the Pearson correlation coefficient $r$ for the correlation between the analyzed property and the specific sky direction, while the column on the right presents the $p-$value of this correlation. From top to bottom, we present the rescaled time at the end of the plateau $T_a^*$, the redshift of the sample distribution and the logarithm of the corrected flux $F_{a_{cor}}$. The turquoise diamond marker represents the best-fit value for the dipole direction obtained in the Bayesian MCMC analysis.}
\label{fig:individual_properties}
\end{figure*}

\section{Dipolar analysis of individual GRB properties}

Based on previous works in which the radiative properties of GRB emission have been studied \citep{Ripa:2017scm, Ripa:2018hak, Lopes:2024cnw}, we have included in this Appendix the results obtained by calculating the full-sky correlation between the radiative property and the angle between the GRB source and the test direction. The radiative properties analyzed are the  rescaled time at the end of the plateau $T_a^*$, the redshift of the sample distribution and the logarithm of the corrected flux $F_{a_{cor}}$.
Note that besides the low correlation values, the directions are always consistent with each other and consistent with the dipole best-fit Bayesian MCMC analysis result. Note particularly how the $p-$values obtained for the corrected flux achieves values below $0.001$, reinforcing the comprehension that these alignments and correlations are not obtained by chance.

\section{Connecting the dipole with a boost velocity}
\label{A: boost}
One of the main caveats of the Dipole fitting method \citep{Mariano:2012wx} here adopted is the lack of an easy straightforward interpretation of the dipole signal in terms of a boost velocity, making it a bit unclear to answer what is the velocity we are moving with respect to the GRB rest-frame background. For this, we have decided to include below the derivation of how to go from one to the other. Starting from the definition of the distance modulus in terms of the luminosity distance, an observer that is moving with a boost velocity $v^a$ with respect to the GRB rest-frame will measure:
\begin{equation}
    \mu(z) = 5\log_{10} \tilde{d}_L(z) +25\;,
    \label{eq:muz}
\end{equation}
where up to first order in $v$
\begin{equation}
    \tilde{d}_L(z) = (1 -v^a n_a) d_L(z_{m})
\end{equation}
represents the luminosity distance as seen by the boosted observer (in our case, the heliocentric frame) and $z_m$ stands for the GRB redshift as measured by the observer who is at rest in the reference frame that sees no dipole in the GRB matter distribution:
\begin{equation}
    (1 + z) = (1+z_m)(1-v^an_a)\;.
\end{equation}
With this in mind, equation \eqref{eq:muz} becomes:
\begin{equation}
    \mu(z) = 5\log_{10} d_L(z_m) +5\log_{10}(1-v^an_a) +25\;.
\end{equation}
Let us now focus on expanding the first term on the right-hand-side:
\begin{equation}
    5\log_{10} d_L(z_m) = 5\log_{10}\left[ \frac{(1+z_m)c}{H_0}\int_0^{z_m}\frac{dz'}{E(z')}\right]\;,
\end{equation}
where $E(z) = \sqrt{\Omega_r(1+z)^4 + \Omega_m(1+z)^3 +\Omega_k(1+z)^2 +\Omega_{\Lambda}}$. Expanding first order in $v^a$, we have:
\begin{strip}
\begin{align}
    5\log_{10} d_L(z_m) &= 5\log_{10}\left[ \frac{(1+z)(1-v^an_a)c}{H_0}\int_0^{z -v^an_a(1+z)}\frac{dz'}{E(z')}\right]\\
    &=5\log_{10}\left[ \frac{(1+z)(1-v^an_a)c}{H_0}\left(\int_0^{z}\frac{dz'}{E(z')} - \int_{z -v^an_a(1+z)}^z\frac{dz'}{E(z')}\right)\right]\;.
\end{align}
\end{strip}

\begin{strip}
Using the approximation $\int_x^{x+\Delta} f(x')dx' = f(x)\Delta$ for $\Delta\ll1$, we have:
\begin{align}
    5\log_{10} d_L(z_m) &= 5\log_{10} \left[ d_L(z)(1-v^an_a) -\frac{c}{H_0}\frac{v^an_a(1+z)^2}{E(z)}
    \right]\\
    &= 5\log_{10}d_L(z) +5\log_{10} \left[1 -v^an_a\left( 1 +\frac{c}{H_0}\frac{(1+z)^2}{E(z)d_L(z)}\right)\right]\;.
\end{align}
Applying this result to equation \eqref{eq:muz}:
\begin{align}
    \mu(z) &= 5\log_{10}d_L(z) +5\log_{10} \left[1 -v^an_a\left( 1 +\frac{c}{H_0}\frac{(1+z)^2}{E(z)d_L(z)}\right)\right] +5\log_{10} \left(1 -v^an_a\right) +25\\
    &= \mu_{\Lambda CDM}(z) +5\log_{10} \left[1 -v^an_a\left(2 + \frac{c}{H_0}\frac{(1+z)^2}{E(z)d_L(z)}\right) \right]\;.
\end{align}
\end{strip}

Comparing with the dipole fitting method distance modulus, we have:
\begin{equation}
v^an_a = \frac{1-10^{\frac{\mathbf{A}\cdot\mathbf{\hat{n}}}{5}}}{2+\frac{c}{H_0}\frac{(1+z)^2}{E(z)d_L(z)}}\;.
\end{equation}
This indicates that a positive dipole signal in one direction implies that the observer is moving in the \emph{opposite direction}. 

In order to obtain a velocity estimation, one must evaluate the redshift-dependent term.
By plotting it in the redshift range $z\in[0.01,5]$ with $\Omega_m=0.334$, we see that its value is larger for low redshifts (of order 10), with a decay of the form $1/z$ for low $z$ and $1/\log(z)$ for high $z$, indicating the importance of allowing scale variance when doing the dipole fit. Taking it's average value $(1.042)$ for the GRB combined sample aiming to obtain an order of magnitude for the velocity, for the Combined sample we have $A=0.6$, giving us a velocity estimation that is $86\pm 32$ times the CMB velocity. 
This amplitude estimation, however, must be taken with only as an order of magnitude estimation at best since, as mentioned, the theoretical term actually presents a decay in redshift, which must be incorporated in the analysis. This will be left for a future work.


\end{document}